\documentclass{acmtrans2e}

\usepackage{eepic}
\usepackage{epic}
\usepackage{latexsym}
\usepackage{alltt}

\newtheorem{theorem}{Theorem}[section]

\newtheorem{corollary}[theorem]{Corollary}

\newtheorem{lemma}[theorem]{Lemma}
\newtheorem{note}[theorem]{Note}
\newdef{definition}[theorem]{Definition}
\newdef{example}[theorem]{Example}
\newdef{remark}[theorem]{Remark}

\newcommand{\eclipse}{ECL$^i$PS$^e$}

\def\smallromani{\renewcommand{\theenumi}{\roman{enumi}}
\renewcommand{\labelenumi}{(\theenumi)}}

\newcommand{\ES}{\mbox{$\emptyset$}}

\newcommand{\A}{\mbox{$\ \wedge\ $}}

\newcommand{\sse}{\mbox{$\:\subseteq\:$}}

\newcommand{\po}{\mbox{$\ \sqsubseteq\ $}}

\newcommand{\fa}{\mbox{$\forall$}}
\newcommand{\te}{\mbox{$\exists$}}

\newcommand{\LL}{\mbox{$\ldots$}}

\newcommand{\C}[1]{\mbox{$\{{#1}\}$}}           

\newcommand{\NI}{\noindent}
\newcommand{\HB}{\hfill{$\Box$}}

\newcommand{\III}{\vspace{3 mm}}
\newcommand{\II}{\vspace{2 mm}}

\newcommand{\szkew}[1]{\relax \setbox0=\hbox{\kern -24pt $\displaystyle#1$\kern 0pt }%
\box0}
{\catcode`\@=11 \global\let\ifjusthvtest@=\iffalse}

\newcounter{oldmycaption}

\newcommand{\p}[2]{\langle #1 \ ; \ #2 \rangle}

\newcommand{\noprint}[1]{}
\newcommand{\noprintpb}[1]{}

\title{The Role of Commutativity in Constraint Propagation Algorithms}

\author{KRZYSZTOF R. APT\\
CWI and University of Amsterdam}

\begin{document}

\begin{abstract}
Constraint propagation algorithms form an important part of most
of the constraint programming systems.
We provide here a simple, yet very general framework that allows us to
explain several constraint propagation algorithms in a systematic way.
In this framework we proceed in two steps.
First, we introduce a  generic iteration algorithm on partial
orderings and prove its correctness in an abstract setting. Then we
instantiate this algorithm with specific partial orderings and
functions to obtain specific constraint propagation algorithms.

In particular, using the notions commutativity and semi-commutativity,
we show that the {\tt AC-3}, {\tt PC-2}, {\tt DAC} and {\tt DPC}
algorithms for achieving (directional) arc consistency and
(directional) path consistency are instances of a single generic
algorithm.  The work reported here extends and simplifies that of Apt
\citeyear{Apt99b}.  
\end{abstract}


            


\makeatletter
\renewcommand{\@opargbegintheorem}[3]{\trivlist
        \item[\hskip 10pt \hskip 
\labelsep{\sc{#1}\savebox\@tempboxa{\sc{#3}}\ifdim 
        \wd\@tempboxa>\z@ \hskip 5pt\relax \sc{#2}  \box\@tempboxa\fi.}]
\itshape}

\makeatother

\begin{bottomstuff}
This is a full, revised and corrected version of our
article Apt \citeyear{Apt99c}. \newline
Author's address: 
CWI, P.O. Box 94079, 1090 GB Amsterdam, the Netherlands.

\permission
\copyright\ 2001 ACM \$5.00
\end{bottomstuff}

\maketitle

\section{Introduction}

\subsection{Motivation}
\label{subsec:motivation}

A constraint satisfaction problem, in short CSP, is a finite
collection of relations (constraints), each on some variables.  A
solution to a CSP is an assignment of values to all variables that
satisfies all constraints.  Constraint programming in a nutshell
consists of generating and solving CSP's means of general or domain
specific methods.

This approach to programming became very popular in the eighties and
led to a creation of several new programming languages and systems.
Some of the more known examples include a constraint logic programming
system \eclipse{} (see Aggoun et al.  \citeyear{Ecl95}), a
multi-paradigm programming language Oz (see, e.g., Smolka
\citeyear{Smo95}), and the ILOG Solver that is the core C++ library
of the ILOG Optimization Suite (see \citeN{ILOG98}).

One of the most important general purpose techniques developed in this
area is constraint propagation that aims at reducing the search space
of the considered CSP while maintaining equivalence.  It is a very
widely used concept. For instance on Google,
\verb+http://www.google.com/+ on December 6th, 2000 the query
``constraint propagation'' yielded 26800 hits. For comparison, the
query ``NP completeness'' yielded 21000 hits.  In addition, in the
literature several other names have been used for the constraint
propagation algorithms: consistency, local consistency, consistency
enforcing, Waltz, filtering or narrowing algorithms.

The constraint propagation algorithms usually aim at reaching some
form of ``local consistency'', a notion that in a loose sense
approximates the notion of ``global consistency''.  Over the last
twenty few years many useful notions of local consistency were
identified and for each of them one or more constraint propagation
algorithms were proposed.

Many of these algorithms were built into the existing constraint
programming systems, including the above three ones. These algorithms
can be triggered either automatically, for example each time a new
constraint is generated (added to the ``constraint store''), or by
means of specific instructions available to the user.

In Apt \citeyear{Apt99b} we introduced a simple framework that allowed
us to explain many of these algorithms in a uniform way.  In this
framework the notion of chaotic iterations, so fair iterations of
functions, on Cartesian products of specific partial orderings played
a crucial role.  We stated there that ``the attempts of finding
general principles behind the constraint propagation algorithms
repeatedly reoccur in the literature on constraint satisfaction
problems spanning the last twenty years'' and devoted three pages to
survey this work.  Two references that are perhaps closest to our work
are Benhamou \citeyear{Ben96} and Telerman and Ushakov
\citeyear{TU96}.

These developments led to an identification of a number of
mathematical properties that are of relevance for the considered
functions, namely monotonicity, inflationarity and idempotence (see,
e.g., Saraswat, Rinard and Panangaden \citeyear{saraswat-semantic} and
Benhamou and Older \citeyear{BO97}). Functions that satisfy these
properties are called closures (see Gierz et al. \citeyear{GHKLMS80}).
Here we show that also the notions of
commutativity and so-called semi-commutativity are important.

As in Apt \citeyear{Apt99b}, to explain the constraint propagation
algorithms, we proceed here in two steps.  First, we introduce a
generic iteration algorithm that aims to compute the least common
fixpoint of a set of functions on a partial ordering and prove its
correctness in an abstract setting.  Then we instantiate this
algorithm with specific partial orderings and functions.  The partial
orderings will be related to the considered variable domains and the
assumed constraints, while the functions will be the ones that
characterize considered notions of local consistency in terms of
fixpoints.

This presentation allows us to clarify which properties of the
considered functions are responsible for specific properties of the
corresponding algorithms.  The resulting analysis is simpler than that
of Apt \citeyear{Apt99b} because we concentrate here on constraint
propagation algorithms that always terminate. This allows us to
dispense with the notion of fairness.  Moreover, we prove here
stronger results by taking into account the commutativity and
semi-commutativity information.

\subsection{Example}
\label{subsec:example}

To illustrate the problems here studied consider the following
puzzle from Mackworth \citeyear{mackworth-constraint-encylopedia}.
\begin{figure}
\begin{center}
\setlength{\unitlength}{0.00041667in}
\begingroup\makeatletter\ifx\SetFigFont\undefined
\def\x#1#2#3#4#5#6#7\relax{\def\x{#1#2#3#4#5#6}}%
\expandafter\x\fmtname xxxxxx\relax \def\y{splain}%
\ifx\x\y   
\gdef\SetFigFont#1#2#3{%
  \ifnum #1<17\tiny\else \ifnum #1<20\small\else
  \ifnum #1<24\normalsize\else \ifnum #1<29\large\else
  \ifnum #1<34\Large\else \ifnum #1<41\LARGE\else
     \huge\fi\fi\fi\fi\fi\fi
  \csname #3\endcsname}%
\else
\gdef\SetFigFont#1#2#3{\begingroup
  \count@#1\relax \ifnum 25<\count@\count@25\fi
  \def\x{\endgroup\@setsize\SetFigFont{#2pt}}%
  \expandafter\x
    \csname \romannumeral\the\count@ pt\expandafter\endcsname
    \csname @\romannumeral\the\count@ pt\endcsname
  \csname #3\endcsname}%
\fi
\fi\endgroup
{\renewcommand{\dashlinestretch}{30}
\begin{picture}(3024,3639)(0,-10)
\path(1212,1212)(1812,1212)
\path(1212,1812)(1212,1212)
\path(12,1812)(612,1812)(612,1212)
\blacken\path(612,3012)(1212,3012)(1212,2412)
	(612,2412)(612,3012)
\path(612,3012)(1212,3012)(1212,2412)
	(612,2412)(612,3012)
\blacken\path(1812,3012)(2412,3012)(2412,2412)
	(1812,2412)(1812,3012)
\path(1812,3012)(2412,3012)(2412,2412)
	(1812,2412)(1812,3012)
\path(12,1812)(12,1212)
\path(12,1212)(12,612)
\path(1212,3612)(1212,3012)
\path(12,3612)(12,3012)
\path(612,2412)(612,1812)
\path(1212,1812)(1812,1812)(1812,1212)
\path(612,2412)(1212,2412)(1212,1812)
\blacken\path(2412,612)(3012,612)(3012,12)
	(2412,12)(2412,612)
\path(2412,612)(3012,612)(3012,12)
	(2412,12)(2412,612)
\put(2412,3462){\makebox(0,0)[lb]{\smash{{{\SetFigFont{6}{7.2}{rm}3}}}}}
\put(612,2262){\makebox(0,0)[lb]{\smash{{{\SetFigFont{6}{7.2}{rm}4}}}}}
\put(1812,2262){\makebox(0,0)[lb]{\smash{{{\SetFigFont{6}{7.2}{rm}5}}}}}
\put(12,1662){\makebox(0,0)[lb]{\smash{{{\SetFigFont{6}{7.2}{rm}6}}}}}
\put(1212,1662){\makebox(0,0)[lb]{\smash{{{\SetFigFont{6}{7.2}{rm}7}}}}}
\put(12,1062){\makebox(0,0)[lb]{\smash{{{\SetFigFont{6}{7.2}{rm}8}}}}}
\put(1212,3462){\makebox(0,0)[lb]{\smash{{{\SetFigFont{6}{7.2}{rm}2}}}}}
\blacken\path(1212,612)(1812,612)(1812,12)
	(1212,12)(1212,612)
\path(1212,612)(1812,612)(1812,12)
	(1212,12)(1212,612)
\blacken\path(612,612)(1212,612)(1212,12)
	(612,12)(612,612)
\path(612,612)(1212,612)(1212,12)
	(612,12)(612,612)
\blacken\path(12,3012)(612,3012)(612,2412)
	(12,2412)(12,3012)
\path(12,3012)(612,3012)(612,2412)
	(12,2412)(12,3012)
\blacken\path(12,2412)(612,2412)(612,1812)
	(12,1812)(12,2412)
\path(12,2412)(612,2412)(612,1812)
	(12,1812)(12,2412)
\path(612,612)(612,1212)
\path(612,612)(612,1212)
\put(12,3462){\makebox(0,0)[lb]{\smash{{{\SetFigFont{6}{7.2}{rm}1}}}}}
\path(12,612)(612,612)
\path(2412,2412)(3012,2412)(3012,1812)
	(2412,1812)(2412,2412)
\path(1812,1812)(2412,1812)(2412,1212)
	(1812,1212)(1812,1812)
\path(2412,1812)(3012,1812)(3012,1212)
	(2412,1212)(2412,1812)
\path(2412,1212)(3012,1212)(3012,612)
	(2412,612)(2412,1212)
\path(1812,1212)(2412,1212)(2412,612)
	(1812,612)(1812,1212)
\path(1212,1212)(1812,1212)(1812,612)
	(1212,612)(1212,1212)
\path(1212,2412)(1812,2412)(1812,1812)
	(1212,1812)(1212,2412)
\path(612,3612)(1212,3612)(1212,3012)
	(612,3012)(612,3612)
\path(1812,3612)(2412,3612)(2412,3012)
	(1812,3012)(1812,3612)
\path(612,3612)(1212,3612)(1212,3012)
	(612,3012)(612,3612)
\blacken\path(612,1812)(1212,1812)(1212,1212)
	(612,1212)(612,1737)
\path(612,1812)(1212,1812)(1212,1212)
	(612,1212)(612,1737)
\path(1212,3012)(1812,3012)(1812,2412)
	(1212,2412)(1212,3012)
\path(2412,3012)(3012,3012)(3012,2412)
	(2412,2412)(2412,3012)
\path(1812,612)(2412,612)(2412,12)
	(1812,12)(1812,612)
\path(2412,3612)(2412,3012)
\path(612,1812)(1212,1812)
\path(1812,1812)(2412,1812)
\path(1812,2412)(2412,2412)(2412,1812)
\path(1812,2412)(1812,1812)
\path(12,1212)(612,1212)
\path(2412,3612)(3012,3612)(3012,3012)
\path(12,612)(612,612)(612,12)
	(12,12)(12,612)
\path(12,3012)(612,3012)
\path(12,3612)(612,3612)(612,3012)
\path(1212,3012)(1812,3012)
\path(1212,3612)(1812,3612)(1812,3012)
\path(2412,3012)(3012,3012)
\end{picture}
}
\caption{A crossword grid \label{fig:crossword}}
\end{center}
\end{figure}
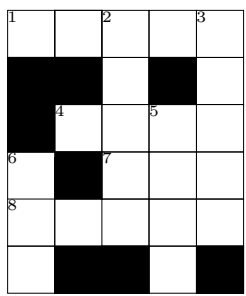
Take the crossword grid of Figure \ref{fig:crossword}
and suppose that we are to fill it with the words
from the following list:

\begin{itemize}
\item HOSES, LASER, SAILS, SHEET, STEER,

\item HEEL, HIKE, KEEL, KNOT, LINE,

\item AFT, ALE, EEL, LEE, TIE.
\end{itemize}
This problem has a unique solution depicted in
Figure \ref{fig:crossword1}.

\begin{figure}
\begin{center}
\setlength{\unitlength}{0.00041667in}
\begingroup\makeatletter\ifx\SetFigFont\undefined
\def\x#1#2#3#4#5#6#7\relax{\def\x{#1#2#3#4#5#6}}%
\expandafter\x\fmtname xxxxxx\relax \def\y{splain}%
\ifx\x\y   
\gdef\SetFigFont#1#2#3{%
  \ifnum #1<17\tiny\else \ifnum #1<20\small\else
  \ifnum #1<24\normalsize\else \ifnum #1<29\large\else
  \ifnum #1<34\Large\else \ifnum #1<41\LARGE\else
     \huge\fi\fi\fi\fi\fi\fi
  \csname #3\endcsname}%
\else
\gdef\SetFigFont#1#2#3{\begingroup
  \count@#1\relax \ifnum 25<\count@\count@25\fi
  \def\x{\endgroup\@setsize\SetFigFont{#2pt}}%
  \expandafter\x
    \csname \romannumeral\the\count@ pt\expandafter\endcsname
    \csname @\romannumeral\the\count@ pt\endcsname
  \csname #3\endcsname}%
\fi
\fi\endgroup
{\renewcommand{\dashlinestretch}{30}
\begin{picture}(3024,3639)(0,-10)
\put(612,2262){\makebox(0,0)[lb]{\smash{{{\SetFigFont{6}{7.2}{rm}4}}}}}
\put(1812,2262){\makebox(0,0)[lb]{\smash{{{\SetFigFont{6}{7.2}{rm}5}}}}}
\put(12,1662){\makebox(0,0)[lb]{\smash{{{\SetFigFont{6}{7.2}{rm}6}}}}}
\put(1212,1662){\makebox(0,0)[lb]{\smash{{{\SetFigFont{6}{7.2}{rm}7}}}}}
\put(1437,3237){\makebox(0,0)[lb]{\smash{{{\SetFigFont{10}{12.0}{rm}S}}}}}
\put(837,3237){\makebox(0,0)[lb]{\smash{{{\SetFigFont{10}{12.0}{rm}O}}}}}
\put(237,3237){\makebox(0,0)[lb]{\smash{{{\SetFigFont{10}{12.0}{rm}H}}}}}
\put(12,1062){\makebox(0,0)[lb]{\smash{{{\SetFigFont{6}{7.2}{rm}8}}}}}
\put(2412,3462){\makebox(0,0)[lb]{\smash{{{\SetFigFont{6}{7.2}{rm}3}}}}}
\blacken\path(612,612)(1212,612)(1212,12)
	(612,12)(612,612)
\path(612,612)(1212,612)(1212,12)
	(612,12)(612,612)
\blacken\path(1212,612)(1812,612)(1812,12)
	(1212,12)(1212,612)
\path(1212,612)(1812,612)(1812,12)
	(1212,12)(1212,612)
\blacken\path(2412,612)(3012,612)(3012,12)
	(2412,12)(2412,612)
\path(2412,612)(3012,612)(3012,12)
	(2412,12)(2412,612)
\blacken\path(1812,3012)(2412,3012)(2412,2412)
	(1812,2412)(1812,3012)
\path(1812,3012)(2412,3012)(2412,2412)
	(1812,2412)(1812,3012)
\blacken\path(12,3012)(612,3012)(612,2412)
	(12,2412)(12,3012)
\path(12,3012)(612,3012)(612,2412)
	(12,2412)(12,3012)
\put(1212,3462){\makebox(0,0)[lb]{\smash{{{\SetFigFont{6}{7.2}{rm}2}}}}}
\put(12,3462){\makebox(0,0)[lb]{\smash{{{\SetFigFont{6}{7.2}{rm}1}}}}}
\path(612,612)(612,1212)
\path(612,612)(612,1212)
\blacken\path(12,2412)(612,2412)(612,1812)
	(12,1812)(12,2412)
\path(12,2412)(612,2412)(612,1812)
	(12,1812)(12,2412)
\put(2037,3237){\makebox(0,0)[lb]{\smash{{{\SetFigFont{10}{12.0}{rm}E}}}}}
\put(2637,837){\makebox(0,0)[lb]{\smash{{{\SetFigFont{10}{12.0}{rm}R}}}}}
\put(2037,2037){\makebox(0,0)[lb]{\smash{{{\SetFigFont{10}{12.0}{rm}K}}}}}
\put(2037,237){\makebox(0,0)[lb]{\smash{{{\SetFigFont{10}{12.0}{rm}L}}}}}
\put(1437,1437){\makebox(0,0)[lb]{\smash{{{\SetFigFont{10}{12.0}{rm}L}}}}}
\put(2637,2637){\makebox(0,0)[lb]{\smash{{{\SetFigFont{10}{12.0}{rm}T}}}}}
\put(837,2037){\makebox(0,0)[lb]{\smash{{{\SetFigFont{10}{12.0}{rm}H}}}}}
\put(837,837){\makebox(0,0)[lb]{\smash{{{\SetFigFont{10}{12.0}{rm}A}}}}}
\put(237,837){\makebox(0,0)[lb]{\smash{{{\SetFigFont{10}{12.0}{rm}L}}}}}
\put(1437,2037){\makebox(0,0)[lb]{\smash{{{\SetFigFont{10}{12.0}{rm}I}}}}}
\put(2637,1437){\makebox(0,0)[lb]{\smash{{{\SetFigFont{10}{12.0}{rm}E}}}}}
\put(2637,2037){\makebox(0,0)[lb]{\smash{{{\SetFigFont{10}{12.0}{rm}E}}}}}
\put(1437,837){\makebox(0,0)[lb]{\smash{{{\SetFigFont{10}{12.0}{rm}S}}}}}
\put(2637,3237){\makebox(0,0)[lb]{\smash{{{\SetFigFont{10}{12.0}{rm}S}}}}}
\put(2037,1437){\makebox(0,0)[lb]{\smash{{{\SetFigFont{10}{12.0}{rm}E}}}}}
\put(1437,2637){\makebox(0,0)[lb]{\smash{{{\SetFigFont{10}{12.0}{rm}A}}}}}
\put(237,1437){\makebox(0,0)[lb]{\smash{{{\SetFigFont{10}{12.0}{rm}A}}}}}
\put(237,237){\makebox(0,0)[lb]{\smash{{{\SetFigFont{10}{12.0}{rm}E}}}}}
\put(2037,837){\makebox(0,0)[lb]{\smash{{{\SetFigFont{10}{12.0}{rm}E}}}}}
\blacken\path(612,3012)(1212,3012)(1212,2412)
	(612,2412)(612,3012)
\path(612,3012)(1212,3012)(1212,2412)
	(612,2412)(612,3012)
\path(2412,1212)(3012,1212)(3012,612)
	(2412,612)(2412,1212)
\path(1812,1212)(2412,1212)(2412,612)
	(1812,612)(1812,1212)
\path(1212,1212)(1812,1212)(1812,612)
	(1212,612)(1212,1212)
\path(1812,612)(2412,612)(2412,12)
	(1812,12)(1812,612)
\path(1212,3012)(1812,3012)
\path(12,3612)(612,3612)(612,3012)
\path(12,3012)(612,3012)
\path(12,612)(612,612)(612,12)
	(12,12)(12,612)
\path(2412,1812)(3012,1812)(3012,1212)
	(2412,1212)(2412,1812)
\blacken\path(612,1812)(1212,1812)(1212,1212)
	(612,1212)(612,1737)
\path(612,1812)(1212,1812)(1212,1212)
	(612,1212)(612,1737)
\path(612,3612)(1212,3612)(1212,3012)
	(612,3012)(612,3612)
\path(1812,3612)(2412,3612)(2412,3012)
	(1812,3012)(1812,3612)
\path(612,3612)(1212,3612)(1212,3012)
	(612,3012)(612,3612)
\path(1212,3012)(1812,3012)(1812,2412)
	(1212,2412)(1212,3012)
\path(1812,1812)(2412,1812)(2412,1212)
	(1812,1212)(1812,1812)
\path(2412,2412)(3012,2412)(3012,1812)
	(2412,1812)(2412,2412)
\path(1212,2412)(1812,2412)(1812,1812)
	(1212,1812)(1212,2412)
\path(2412,3012)(3012,3012)(3012,2412)
	(2412,2412)(2412,3012)
\path(1212,3612)(1812,3612)(1812,3012)
\path(1212,1812)(1812,1812)(1812,1212)
\path(612,2412)(612,1812)
\path(12,3612)(12,3012)
\path(1212,3612)(1212,3012)
\path(612,2412)(1212,2412)(1212,1812)
\path(12,1812)(612,1812)(612,1212)
\path(1212,1812)(1212,1212)
\path(1212,1212)(1812,1212)
\path(12,1812)(12,1212)
\path(12,1212)(12,612)
\path(612,1812)(1212,1812)
\path(2412,3612)(2412,3012)
\path(2412,3612)(3012,3612)(3012,3012)
\path(2412,3012)(3012,3012)
\path(1812,1812)(2412,1812)
\path(12,612)(612,612)
\path(12,1212)(612,1212)
\path(1812,2412)(1812,1812)
\path(1812,2412)(2412,2412)(2412,1812)
\end{picture}
}
\caption{A solution to the crossword puzzle \label{fig:crossword1}}
\end{center}
\end{figure}

This puzzle can be solved by systematically considering each crossing
and eliminating the words that cannot be used.  Consider for example
the crossing of the positions 2 and 4, in short (2,4). Neither word
HOSES nor LASER can be used in position 2 because no four letter word
(for position 4) exists with S as the second letter.  Similarly, by
considering the crossing (2,8) we deduce that none of the words LASER,
SHEET and STEER can be used in position 2.

The question now is what ``systematically'' means. For example, after
considering the crossings (2,4) and (2,8) should we reconsider the
crossing (2,4)? Our approach clarifies that the answer is ``No''
because the corresponding functions $f_{2,4}$ and $f_{2,8}$ that
remove impossible words, here for position 2 on account of the crossings
(2,4) and (2,8), commute. In contrast, the functions $f_{2,4}$ and
$f_{4,5}$ do not commute, so after considering the crossing (4,5) the
crossing (2,4) needs to be reconsidered.

In Section \ref{sec:ac3}
we formulate this puzzle as a CSP and discuss more precisely
the problem of scheduling of the involved functions and the role
commutativity plays here.

\subsection{Plan of the Paper}
\label{subsec:plan}

This article is organized as follows. First, in Section
\ref{sec:gen-ite}, drawing on the approach of Monfroy and R{\'{e}}ty
\citeyear{MR99}, we introduce a generic iteration algorithm, with the
difference that the partial ordering is not further analyzed. Next, in
Section \ref{sec:compound}, we refine it for the case when the partial
ordering is a Cartesian product of component partial orderings and in
Section \ref{sec:from-to} explain how the introduced notions should be
related to the constraint satisfaction problems.  These last two
sections essentially follow Apt \citeyear{Apt99b}, but because we
started here with the generic iteration algorithms on arbitrary
partial orderings we built now a framework in which we can also discuss the
role of commutativity.

In the next four sections we instantiate the algorithm of
Section \ref{sec:gen-ite} or some of its refinements
to obtain specific constraint propagation
algorithms.  In particular, in Section 
\ref{sec:hyper-arc-algo} we derive
algorithms for arc consistency and hyper-arc
consistency.  These algorithms can be improved by taking into account
information on commutativity. This is done in Section \ref{sec:ac3}
and yields the well-known {\tt AC-3} algorithm.
Next, in Section \ref{sec:path-algo} we derive an algorithm for path
consistency and in Section \ref{sec:pc2} we improve it, again by
using information on commutativity. This yields 
the {\tt PC-2} algorithm.

In Section \ref{sec:simple-ite} we clarify under what assumptions the
generic algorithm of Section \ref{sec:gen-ite} can be simplified to a
simple {\bf for} loop statement.  Then we instantiate this simplified
algorithm to derive in Section \ref{sec:directional-arc-algo} the {\tt
  DAC} algorithm for directional arc consistency and in Section
\ref{sec:directional-path-algo} the {\tt DPC} algorithm for
directional path consistency.  Finally, in Section
\ref{sec:conclusions} we draw conclusions and discuss recent and
possible future work.

We deal here only with the classic algorithms that establish
(directional) arc consistency and (directional) path consistency and
that are more than twenty, respectively ten, years old.  However,
several more ``modern'' constraint propagation algorithms can also be
explained in this framework. In particular, in Apt \citeyear[page
203]{Apt99b} we derived from a generic algorithm a simple algorithm
that achieves the notion of relational consistency of Dechter and van
Beek \citeyear{DvB97}.  In turn, by mimicking the development of Sections
\ref{sec:directional-arc-algo} and \ref{sec:directional-path-algo}, we
can use the framework of Section \ref{sec:simple-ite} to derive the
adaptive consistency algorithm of Dechter and Pearl \citeyear{dechter88}.
Now, Dechter \citeyear{D99} showed that the latter algorithm can be formulated
in a very general framework of bucket elimination that in turn can be
used to explain such well-known algorithms as directional resolution,
Fourier-Motzkin elimination, Gaussian elimination, and also various
algorithms that deal with belief networks.

\section{Generic Iteration Algorithms}
\label{sec:gen-ite}

Our presentation is completely general.  Consequently, we delay the
discussion of constraint satisfaction problems till Section
\ref{sec:from-to}.  In what follows we shall rely on the following
concepts.

\begin{definition}
  Consider a partial ordering $(D, \po )$ with the least element $\bot$
  and a finite set of functions $F := \C{f_1, \LL , f_k}$ on $D$.
 \begin{itemize}

\item 
By an {\em iteration of $F$\/} 
we mean an infinite sequence of values 
$d_0, d_1, \LL  $ defined inductively by
\[
d_0 := \bot,
\]
\[
d_{j} := f_{i_{j}}(d_{j-1}),
\]
where each $i_j$ is an element of $[1..k]$.

\item We say that an increasing sequence
$d_0 \: \po \: d_1 \: \po \: d_2 \: \LL$ of elements from $D$
{\em eventually stabilizes at d\/} if for some $j \geq 0$ we have
$d_i = d$ for $i \geq j$.
\HB
\end{itemize}
\end{definition}

In what follows we shall consider iterations of functions that
satisfy some specific properties.

\begin{definition}
Consider a partial ordering $(D, \po)$ and a function $f$ on $D$.

\begin{itemize}
\item $f$ is called {\em inflationary\/} \index{function!inflationary}
if $x \po f(x)$ for all $x$.

\item $f$ is called {\em monotonic\/} \index{function!monotonic}
if $x \po y$ implies 
$f(x) \po f(y)$ for all $x, y$.
\HB
\end{itemize}
\end{definition}

The following simple observation clarifies the role of monotonicity.
The subsequent result will clarify the role of inflationarity.

\begin{lemma}[(Stabilization)] \label{lem:stabilization}
Consider a partial ordering  $(D, \po )$ with the 
least element $\bot$
and a finite set of  monotonic functions  $F$ on $D$.

Suppose that an
iteration of $F$ eventually stabilizes at a common fixpoint $d$
of the functions from $F$.  Then $d$ is the least common fixed point
of the functions from $F$.
\end{lemma}

\begin{proof}
Consider a common fixpoint $e$ of the functions from $F$. We prove
that $d \po e$.  Let $d_0, d_1, \LL$ be the iteration in question.
For some  $j \geq 0$ we have $d_i = d$ for $i \geq j$.

It suffices to prove by induction on $i$ that  $d_i \po e$. 
The claim obviously holds for $i = 0$ since $d_0 = \bot$.
Suppose it holds for some $i \geq 0$.
We have $d_{i+1} = f_j(d_i)$ for some $j \in [1..k]$.

By the monotonicity of $f_j$ and the induction hypothesis we get 
$f_j(d_i) \po f_j(e)$, so $d_{i+1} \po e$ since $e$ is a fixpoint of $f_j$.
\end{proof}

We fix now a partial ordering $(D, \po )$ with the least element
$\bot$ and a finite set of functions $F$ on $D$.  We
are interested in computing the least common fixpoint of the functions
from $F$.  To this end we study the following algorithm 
inspired by a similar, though more complex, algorithm of
Monfroy and R{\'{e}}ty \citeyear{MR99} defined on a
Cartesian product of component partial orderings.
\III

\NI 
{\sc   Generic Iteration Algorithm ({\tt GI})}
\begin{tabbing}
\= $d := \bot$; \\
\> $G := F$; \\ 
\> {\bf while} $G \neq \ES$ {\bf do} \\
\> \qquad choose $g \in G$; \\
\> \qquad $G := G - \C{g}$; \\
\> \qquad $G := G \cup update(G,g,d)$; \\
\> \qquad $d := g(d)$ \\
\> {\bf od} 
\end{tabbing}

where for all $G,g,d$ the set of functions 
$update(G,g,d)$ from $F$ is such that

\begin{description}
\item[{\bf A}] $\C{f \in F - G  \mid f(d) = d \A f(g(d)) \neq g(d)} \sse update(G,g,d)$,

\item[{\bf B}] $g(d) = d$ implies that $update(G,g,d) = \ES$,

\item[{\bf C}] $g(g(d)) \neq g(d)$ implies that $g \in update(G,g,d)$.
\end{description}

The above conditions on $update(G,g,d)$ look somewhat artificial and
unnecessarily complex. In fact, an obviously simpler alternative
exists according to which we just postulate that $\C{f \in F - G \mid
  f(g(d)) \neq g(d)} \sse update(G,g,d)$, i.e., that we add to $G$ at least
all functions from $F - G$ for which the ``new value'', $g(d)$, is not
a fixpoint.

The problem is that for each choice of the $update$ function we wish
to avoid the computationally expensive task of computing the values of
$f(d)$ and $f(g(d))$ for the functions $f$ in $F - G$. Now, when we
specialize the above algorithm to the case of a Cartesian product of
the partial orderings we shall be able to avoid this computation of
the values of $f(d)$ and $f(g(d))$ by just analyzing for which
components $d$ and $g(d)$ differ.  This specialization cannot be
derived by adopting the above simpler choice of the $update$ function.

Intuitively, assumption {\bf A} states that $update(G,g,d)$ at least contains
all the functions from $F - G$ for which the ``old value'', 
$d$, is a fixpoint but the ``new value'', $g(d)$, is not.
So at each loop iteration such functions are added to the set $G$.
In turn, assumption {\bf B} states that no functions are added to $G$
in case the value of $d$ did not change.
Note that even though after the assignment $G := G - \C{g}$ we have
$g \in F - G$, still 
$g \in \C{f \in F - G  \mid f(d) = d \A f(g(d)) \neq g(d)}$ does not 
hold since we cannot have both $g(d) = d$ and $g(g(d)) \neq g(d)$.
So assumption {\bf A} does not provide any information when $g$ is
to be added back to $G$. This information is provided in assumption {\bf C}.

On the whole, the idea is to keep in $G$ at least all functions $f$
for which the current value of $d$ is not a fixpoint.

An obvious example of an $update$ function that satisfies assumptions {\bf A},
{\bf B} and {\bf C} is 
\[
update(G,g,d) := \C{f \in F - G  \mid f(d) = d \A f(g(d)) \neq g(d)} \cup {\bf C}(g),
\]
where
\[
\mbox{${\bf C}(g) = \C{g}$ if $g(g(d)) \neq g(d)$ and otherwise ${\bf C}(g) = \ES $.}
\]
However, again, this choice of the $update$ function is computationally
expensive because for each function $f$ in $F - G$ we would have to
compute the values $f(g(d))$ and $f(d)$. 

We now prove correctness of this algorithm in the following sense.

\begin{theorem}[({\tt GI})] \label{thm:GI}
  \mbox{} \\[-6mm]
\begin{enumerate} \smallromani
\item Every terminating execution of the {\tt GI} algorithm computes
in $d$ a common fixpoint of the functions from $F$.

\item  Suppose that all functions in $F$ are monotonic.
Then every terminating execution of the {\tt GI} algorithm computes
in $d$ the least common fixpoint of the functions from $F$.

\item  Suppose that all functions in $F$ are inflationary and
that $(D, \po)$ is finite. Then every
  execution of the {\tt GI} algorithm terminates.
\end{enumerate}
\end{theorem}

\begin{proof}
\mbox{}

\NI
$(i)$
Consider the predicate $I$ defined by:
\[
I :=\fa f \in F - G \  f(d) = d.
\]
Note that $I$ is established by the assignment $G := F$.  Moreover, it
is easy to check that by virtue of assumptions {\bf A}, {\bf B} and
{\bf C} $I$ is preserved by each {\bf while} loop iteration.  Thus $I$
is an invariant of the {\bf while} loop of the algorithm.  (In fact,
assumptions {\bf A}, {\bf B} and {\bf C} are so chosen that $I$
becomes an invariant.)  Hence upon its termination
\[
(G = \ES) \A I
\]
holds, that is  
\[
\fa f \in F \: f(d) = d.
\]

\II

\NI
$(ii)$
This is a direct consequence of $(i)$  and the
Stabilization Lemma \ref{lem:stabilization}.
\II

\NI
$(iii)$
Consider the lexicographic ordering of the strict partial orderings
$(D, \sqsupset)$ and $({\cal N}, <)$, 
defined on the elements of $D \times {\cal N}$ by
\[ 
(d_1, n_1) <_{lex} (d_2, n_2)\ {\rm iff} \ d_1 \sqsupset d_2
        \ {\rm or}\ ( d_1 = d_2 \ {\rm and}\ n_1 < n_2). 
\]
We use here the inverse ordering $\sqsupset$
defined by: $d_1 \sqsupset d_2$ iff $d_2 \sqsubseteq d_1$ and $d_2 \neq d_1$.

Given a finite set $G$ we denote by $card \ G$ the number of its
elements.  By assump\-tion all functions in $F$ are inflationary so, by
virtue of assumption {\bf B}, with each {\bf while} loop iteration of
the modified algorithm the pair
\[
(d, card \ G)
\]
strictly decreases in this ordering $<_{lex}$.
But by assumption 
$(D, \po )$ is finite, so
$(D, \sqsupset)$ is well-founded and consequently so is
$(D \times {\cal N}, <_{lex})$. This implies termination.
\end{proof}

In particular, we obtain the following conclusion.

\begin{corollary}[({\tt GI})] \label{cor:GI}
  Suppose that $(D, \po )$ is a finite partial ordering with the least
  element $\bot$. Let $F$ be a finite set of monotonic and
  inflationary functions on $D$. Then every execution of the {\tt GI}
  algorithm terminates and computes in $d$ the least common fixpoint
  of the functions from $F$.  
\end{corollary}

In practice, we are not only interested that the {\em update\/} function
is easy to compute but also that it generates small sets of functions.
Therefore we show how the function $update$ can be made smaller when
some additional information about the functions in $F$ is available.
This will yield specialized versions of the {\tt GI} algorithm.  First
we need the following simple concepts.

\begin{definition} Consider two functions $f, g$ on a set $D$.
  \begin{itemize}
  \item 
We say that  $f$ and $g$ {\em commute\/} if 
$f (g (x)) = g (f (x))$ for all $x$.

\item We call $f$ {\em idempotent} if
$f(f(x)) = f(x)$ for all $x$.

\item
We call a function $f$ on a partial ordering $(D, \po)$
a {\em closure\/} if $f$ is inflationary, monotonic and itempotent.
\HB  
\end{itemize}
\end{definition}

Closures were studied in Gierz et al. \citeyear{GHKLMS80}.  They play
an important role in mathematical logic and lattice theory. We shall
return to them in Section \ref{sec:from-to}.

The following result holds.

\begin{theorem}[(Update)] \label{thm:update} 
\mbox{} \\[-6mm]

\begin{enumerate} \smallromani

\item If $update(G,g,d)$ satisfies assumptions {\bf A}, {\bf B} and {\bf C},
then so does the function
\[
update(G,g,d) - Idemp(g),
\]
where
\[
\mbox{$Idemp(g) = \C{g}$ if $g$ is idempotent and otherwise $Idemp(g) = \ES $.}
\]

\item Suppose that for each $g$ the set of functions $Comm(g)$
from $F$ is such that
\begin{itemize}

\item $g \not\in Comm(g)$,

\item each element of $Comm(g)$ commutes with $g$.

\end{itemize}
If $update(G,g,d)$ satisfies assumptions {\bf A}, {\bf B} and {\bf C},
then so does the function
\[
update(G,g,d) - Comm(g).
\]
\end{enumerate}
\end{theorem}

\begin{proof}
It suffices to establish in each case assumption {\bf A} and {\bf C}.
Let 
\[
A := \C{f \in F - G  \mid f(d) = d \A f(g(d)) \neq g(d)}.
\]

\NI
$(i)$
After introducing the {\tt GI} algorithm we noted already that
$g \not \in A$. So assumption {\bf A} implies 
$A \sse  update(G,g,d) - \C{g}$ and a fortiori $A \sse  update(G,g,d) - Idemp(g)$.

For assumption {\bf C} it suffices to note that $g(g(d)) \neq g(d)$ implies
that $g$ is not idempotent, i.e., that $Idemp(g) =  \ES$.
\II

\NI 
$(ii)$ Consider $f \in A$. Suppose that $f \in Comm(g)$. Then $f(g(d) =
g(f(d)) = g(d)$ which is a contradiction. So $f \not\in Comm(g)$.
Consequently, assumption {\bf A} implies $A \sse update(G,g,d) - Comm(g)$.

For assumption {\bf C} it suffices to use the fact that $g \not\in Comm(g)$.
\end{proof}

We conclude that given an instance of the {\tt GI} algorithm that
employs a specific $update$ function, we can obtain other instances of
it by using $update$ functions modified as above.  Note that both
modifications are independent of each other and therefore can be
applied together.

In particular, when each function is idempotent
and the function $Comm$ satisfies the assumptions of $(ii)$, then
the following holds:
if $update(G,g,d)$ satisfies assumptions {\bf A}, {\bf B} and {\bf C},
then so does the function $update(G,g,d) - (Comm(g) \cup \C{g})$.

\section{Compound Domains}
\label{sec:compound}

In the applications we study the iterations are carried out on a
partial ordering that is a Cartesian product of the partial orderings.
So assume now that the partial ordering $(D, \po)$ is
the Cartesian product of some partial orderings $(D_i, \po_i)$, for $i
\in [1..n]$, each with the least element $\bot_i$.  So $D = D_1 \times
\cdots \times D_n$.

 Further, we assume that each
function from $F$ depends from and affects only certain components of
$D$.  To be more precise we introduce a simple notation and
terminology.

\begin{definition} 
  Consider a sequence of partial orderings $(D_1, \po_1), \LL , (D_n,
  \po_n)$.
  \begin{itemize}

  \item By a {\em scheme\/} (on $n$) we mean a growing sequence of
different elements from $[1..n]$.  

  \item Given a scheme $s := i_1,\LL, i_l$ on $n$ we denote by 
  $(D_s, \po_s)$ the Cartesian product of the
  partial orderings $(D_{i_j}, \po_{i_j})$, for $j \in [1..l]$.

  \item Given a function $f$ on $D_s$ we say that $f$ is {\em with
      scheme $s$} and say that $f$ {\em depends on $i$\/} if $i$ is an
    element of $s$.

  \item Given an $n$-tuple $d := d_1, \LL, d_n$ from $D$ and a scheme
    $s := i_1, \LL, i_l$ on $n$ we denote by $d[s]$ the tuple
    $d_{i_1}, \LL , d_{i_l}$.  In particular, for $j \in [1..n]$ \ 
    $d[j]$ is the $j$-th element of $d$.  \HB
\end{itemize}
\end{definition} 

Consider now a function $f$ with scheme $s$.  We
extend it to a function $f^+$ from $D$ to $D$ as follows.
Take  $d \in D$. We set
\[
f^+(d) := e
\]
where  $e[s] = f(d[s])$ and $e[n-s] = d[n-s]$,
and where $n-s$ is the scheme obtained by removing from $1, \LL, n$
the elements of $s$.
We call  $f^+$ the {\em canonic extension\/} of $f$ to the domain $D$.

So $f^+(d_1, \LL, d_n) = (e_1, \LL, e_n)$ implies $d_i = e_i$
for any $i$ not in the scheme $s$ of $f$.
Informally, we can summarize it by saying that $f^+$ does not change the
components on which it does not depend. This is what we meant above by
stating that each considered function affects only certain components
of $D$.

We now say that two functions, $f$ with scheme  $s$
and $g$ with scheme $t$, {\em commute\/} if the functions
$f^+$ and $g^+$ commute.

Instead of defining iterations for the case of the functions with
schemes, we rather reduce the situation to the one studied in the
previous section and consider, equivalently, the iterations of the
canonic extensions of these functions to the common domain $D$.
However, because of this specific form of the considered functions, we
can use now a simple definition of the $update$ function.  More
precisely, we have the following observation.

\begin{note}[(Update)] \label{note:update}
Suppose that each function in $F$ is of the form $f^+$. Then the 
following function $update$ satisfies assumptions {\bf A}, {\bf B} and {\bf C}:
\[
update(G,g^+,d):= \C{f^+ \in F  \mid f \mbox{ depends on some } i \mbox{ in $s$ such that } d[i] \neq  g^+(d)[i]}, 
\]
where $g$ is with scheme $s$.
\end{note}
\begin{proof}
To deal with assumption {\bf A} take a function
$f^+ \in F-G$ such that $f^+(d) = d$. Then 
$f^+(e) = e$ for any $e$ that coincides with $d$ on all components 
that are in the scheme of $f$.

Suppose now additionally that $f^+(g^+(d)) \neq g^+(d)$. By the above
$g^+(d)$ is not such an $e$, i.e., $g^+(d)$ 
differs from $d$ on some component $i$ in the scheme of $f$.
In other words, $f$ depends on some $i$ such that $d[i] \neq
g^+(d)[i]$.  This $i$ is then in the scheme of $g$
and consequently $f^{+} \in update(G,g^+,d)$.

The proof for assumption {\bf B} is immediate.

Finally, to deal with assumption {\bf C} it suffices to note
that $g^+(g^+(d)) \neq g^+(d)$ implies $g^+(d)) \neq d$, which
in turn implies that $g^{+} \in update(G,g^+,d)$.
\end{proof}

This, together with the {\tt GI} algorithm, yields the following
algorithm in which we introduced a variable $d'$ to hold the value of
$g^+(d)$, and used $F_0 := \C{f \mid f^+ \in F}$ and the functions with
schemes instead of their canonic extensions to $D$.  
\II

\NI
{\sc Generic Iteration Algorithm for Compound Domains ({\tt CD})}
\begin{tabbing}
\= $d := (\bot_1, \LL, \bot_n)$; \\
\> $d' := d$; \\ 
\> $G := F_0$; \\ 
\> {\bf while} $G \neq \ES$ {\bf do} \\
\> \qquad choose $g \in G$; suppose $g$ is with scheme $s$; \\
\> \qquad $G := G - \C{g}$; \\
\> \qquad $d'[s] := g(d[s])$; \\
\> \qquad $G := G \cup \C{f \in  F_0  \mid \mbox{$f$ depends on
 some } i \mbox{ in $s$ such that } d[i] \neq d'[i]}$; \\
\> \qquad $d[s] := d'[s]$ \\
\> {\bf od} 
\end{tabbing}

The following corollary to the {\tt GI} Theorem \ref{thm:GI} and the
Update Note \ref{note:update} summarizes the correctness of this
algorithm. It corresponds to Theorem 11 of Apt \citeyear{Apt99b}
where the iteration algorithms were introduced immediately 
on compound domains.

\begin{corollary}[({\tt CD})] \label{cor:CD}
  Suppose that $(D, \po )$ is a finite partial ordering that is a
  Cartesian product of n partial orderings, each with the least element
  $\bot_i$ with $i \in [1..n]$. Let $F$ be a finite set of functions
  on $D$, each of the form $f^+$.

Suppose that all functions in $F$ are monotonic and inflationary. Then
every execution of the {\tt CD} algorithm terminates and computes in
$d$ the least common fixpoint of the functions from $F$.  
\end{corollary}

In the subsequent presentation we shall deal with the following
two modifications of the {\tt CD} algorithm:

\begin{itemize}
\item {\em {\tt CDI} algorithm}. This is the version of the {\tt CD} algorithm
in which all the functions are idempotent and the function $update$
defined in the Update Theorem \ref{thm:update}$(i)$ is used.

\item {\em {\tt CDC} algorithm}. This is the version of the {\tt CD}
  algorithm in which all the functions are idempotent and the combined
  effect of the functions $update$ defined in the Update Theorem
  \ref{thm:update} is used for some function $Comm$.

\end{itemize}

For both algorithms the counterparts of the {\tt CD} Corollary
\ref{cor:CD} hold.

\section{From Partial Orderings to Constraint Satisfaction Problems}
\label{sec:from-to}

We have been so far completely general in our discussion.  Recall that
our aim is to derive various constraint propagation algorithms.  To be
able to apply the results of the previous section we need to relate
various abstract notions that we used there to constraint satisfaction
problems.

This is perhaps the right place to recall the definition
and to fix the notation.
Consider a finite sequence of variables $X := x_1, \LL, x_n$,
where $n \geq 0$, with respective domains ${\cal D} := D_1, \LL, D_n$
associated with them.  So each variable $x_i$ ranges over the domain
$D_i$.  By a {\em constraint} $C$ on $X$ we mean a subset of $D_1
\times \LL \times D_n$.  

By a {\em constraint satisfaction problem}, in short CSP, we mean a
finite sequence of variables $X$ with respective domains ${\cal
  D}$, together with a finite set $\cal C$ of constraints, each on a
subsequence of $X$. We write it as $\p{{\cal C}}{x_1 \in D_1,
  \LL, x_n \in D_n}$, where $X := x_1, \LL, x_n$ and ${\cal D} :=
D_1, \LL, D_n$.

Consider now an element $d := d_1, \LL, d_n$ of $D_1 \times \LL \times
D_n$ and a subsequence $Y := x_{i_1}, \LL, x_{i_\ell}$ of
$X$. Then we denote by $d[Y]$ the sequence
$d_{i_1}, \LL, d_{i_{\ell}}$.

By a {\em solution\/} to  $\p{{\cal C}}{x_1 \in D_1, \LL, x_n \in D_n}$
we mean an element $d \in D_1 \times \LL \times D_n$ such that for
each constraint $C \in {\cal C}$ on a sequence of variables $Y$
we have $d[Y] \in C$.
We call a CSP {\em consistent\/} if it has a solution.  Two CSP's
${\cal P}_1$ and ${\cal P}_2$ with the same sequence of variables are
called {\em equivalent\/} if they have the same set of solutions.
This definition extends in an obvious way to the case of
two CSP's with the same {\em sets\/} of variables.

Let us return now to 
the framework of the previous section. It involved:
\begin{enumerate}\smallromani

\item Partial orderings with the least elements;

These will correspond to partial orderings on the CSP's.
In each of them the original CSP will be the least element
and the partial ordering will be determined by the local
consistency notion we wish to achieve.

\item Monotonic and inflationary functions with schemes;

These will correspond to the functions that transform the variable
domains or the constraints. Each function will be associated with one
or more constraints.

\item Common fixpoints;

These will correspond to the CSP's that satisfy the considered
notion of local consistency. 
\end{enumerate}

Let us be now more specific about items (i) and (ii).
\II

\NI
Re: (i)

To deal with the local consistency notions considered in this paper
we shall introduce two specific partial orderings on the
CSP's. In each of them the considered CSP's will be defined on the
same sequences of variables.

We begin by fixing for each set $D$ a collection ${\cal F}(D)$ of the 
subsets of $D$ that includes $D$ itself. So ${\cal F}$ is a function
that given a set $D$ yields a set of its subsets to which $D$ belongs.

When dealing with the notion of hyper-arc consistency ${\cal F}(D)$ will be simply the
set ${\cal P}(D)$ of all subsets of $D$ but for specific domains only
specific subsets of $D$ will be chosen.  For example, to deal with the
the constraint propagation for the linear constraints on integer
interval domains we need to choose for ${\cal F}(D)$ the set of all
subintervals of the original interval $D$.

When dealing with the path consistency, for a constraint $C$ the
collection ${\cal F}(C)$ will be also the set ${\cal P}(C)$ of all
subsets of $C$. However, in general other choices may be needed.  For
example, to deal with the cutting planes method, we need to limit our
attention to the sets of integer solutions to finite sets of linear
inequalities with integer coefficients (see Apt \citeyear[pages
193-194]{Apt99b}).

Next, given two CSP's, $\phi := \p{{\cal C}}{x_1 \in D_1, \LL, x_n \in D_n}$
and $\psi := \p{{\cal C'}}{x_1 \in D'_1, \LL, x_n \in D'_n}$, we write
$\phi \sqsubseteq_d \psi$ iff 
\begin{itemize}

\item $D'_i \in {\cal F}(D_i)$ (and hence $D'_i \sse D_i$)
for $i \in [1..n]$,

\item the constraints in ${\cal C'}$ are the restrictions of the
  constraints in ${\cal C}$ to the domains $D'_1, \LL, D'_n$.
\end{itemize}

Next, given two CSP's, $\phi := \p{C_1, \LL, C_k}{{\cal DE}}$
and $\psi := \p{C'_1, \LL, C'_k}{{\cal DE}}$, we write
$\phi \sqsubseteq_c \psi$ iff 

\begin{itemize}

\item $C'_i \in {\cal F}(C_i)$ (and hence $C'_i \sse C_i$)
for $i \in [1..k]$.

\end{itemize}

In what follows we call $\sqsubseteq_d$ the {\em domain reduction
  ordering\/} and $\sqsubseteq_c$ the {\em constraint reduction
  ordering}.  To deal with the arc consistency, hyper-arc consistency
and directional arc consistency notions we shall use the domain
reduction ordering, and to deal with path consistency and directional
path consistency notions we shall use the constraint reduction
ordering.

We consider each ordering with some fixed initial CSP ${\cal P}$
as the least element. In other words, each domain reduction ordering
is of the form
\[
(\C{{\cal P'} \mid {\cal P} \sqsubseteq_d {\cal P'}}, \sqsubseteq_d)
\]
and each constraint reduction ordering is of the form
\[
(\C{{\cal P'} \mid {\cal P} \sqsubseteq_c {\cal P'}}, \sqsubseteq_c).
\]
\II

\NI
Re: (ii)

The domain  reduction ordering and the constraint reduction ordering
are not directly amenable to the analysis given in Section \ref{sec:compound}.
Therefore, we shall rather use equivalent partial orderings defined
on compound domains.
To this end note that 
$
\p{{\cal C}}{x_1 \in D'_1, \LL, x_n \in D'_n} \sqsubseteq_d
\p{{\cal C'}}{x_1 \in D''_1, \LL, x_n \in D''_n}$
iff
$D'_i \supseteq D''_i \mbox{ for } i \in [1..n]$.

This equivalence means that for ${\cal P} = \p{{\cal C}}{x_1 \in D_1,
  \LL, x_n \in D_n}$ we can identify the domain reduction ordering
$(\C{{\cal P'} \mid {\cal P} \sqsubseteq_d {\cal P'}}, \sqsubseteq_d)$
with the Cartesian product of the partial orderings $({\cal F}(D_i),
\supseteq)$, where $i \in [1..n]$.

Additionally, each CSP in this domain reduction ordering is uniquely
determined by its domains and by the initial ${\cal P}$.
Indeed, by the definition of this ordering the constraints 
of such a CSP are restrictions of the constraints of ${\cal P}$
to the domains of this CSP.

Similarly, 
\[
\p{C'_1, \LL, C'_k}{{\cal DE}} \sqsubseteq_c  \p{C''_1, \LL, C''_k}{{\cal DE}} \mbox{ iff }
C'_i  \supseteq C''_i \mbox{ for } i \in [1..k].
\]
This allows us for ${\cal P} = \p{C_1, \LL, C_k}{{\cal DE}}$
to identify the constraint reduction ordering
$(\C{{\cal P'} \mid {\cal P} \sqsubseteq_c {\cal P'}}, \sqsubseteq_c)$
with the Cartesian product of the partial orderings 
$({\cal F}(C_i), \supseteq)$, where $i \in [1..k]$.
Also, each CSP in this constraint reduction ordering is uniquely
determined by its constraints and by the initial ${\cal P}$.

In what follows instead of the domain reduction ordering and the
constraint reduction ordering we shall use the corresponding Cartesian
products of the partial orderings.  So in these compound orderings the
sequences of the domains (respectively, of the constraints) are
ordered componentwise by the reversed subset ordering
$\supseteq$. Further, in each component ordering $({\cal F}(D),
\supseteq)$ the set $D$ is the least element.

The reason we use these compound orderings is that we can now employ
functions with schemes, as used in Section \ref{sec:compound}.
Each such function $f$ is defined on a sub-Cartesian product
of the constituent partial orderings. Its canonic extension $f^{+}$,
introduced in Section \ref{sec:compound}, is then defined on the
``whole'' Cartesian product.

Suppose now that we are dealing with the domain reduction ordering 
with the least (initial) CSP ${\cal P}$ and that
\[
f^+(D_1 \times \cdots \times D_n) = D'_1 \times \cdots \times D'_n.
\]
Then the sequence of the domains $(D_1, \LL, D_n)$ and ${\cal P}$
uniquely determine a CSP in this ordering
and the same for $(D'_1, \LL, D'_n)$ and ${\cal P}$.
Hence $f^+$, and a fortiori $f$, can be viewed as a function on
the CSP's that are elements of this domain reduction ordering.
In other words, $f$ can be viewed as a function on CSP's.

The same considerations apply to the constraint reduction ordering.
We shall use these observations when arguing about the equivalence
between the original and the final CSP's for various constraint
propagation algorithms.

The considered functions with schemes will be now used in presence of
the componentwise ordering $\supseteq$. The following observation will
be useful.

Consider a function $f$ on some Cartesian product
${\cal F}(E_1) \times \LL \times {\cal F}(E_m)$.
Note that $f$ is inflationary w.r.t. the componentwise ordering $\supseteq$ if 
for all $(X_1, \LL, X_m) \in {\cal F}(E_1) \times \LL \times {\cal F}(E_m)$
we have $Y_i \subseteq X_i$ for all $i \in [1..m]$, where 
$f(X_1, \LL, X_m) = (Y_1, \LL, Y_m)$.

Also, $f$ is monotonic w.r.t. the componentwise ordering $\supseteq$ if for all 
$(X_1, \LL, X_m),$ $(X'_1, \LL, X'_m) \in {\cal F}(E_1) \times \LL \times {\cal F}(E_m)$
such that $X_i \subseteq X'_i$ for all $i \in [1..m]$,
the following holds: if
\[
\mbox{$f(X_1, \LL, X_m) = (Y_1, \LL, Y_m)$ and
$f(X'_1, \LL, X'_m) = (Y'_1, \LL, Y'_m)$,}
\]
then
$Y_i \subseteq Y'_i$ for all $i \in [1..m]$. 

In other words, $f$ is monotonic w.r.t.
$\supseteq$ iff it is monotonic w.r.t. $\subseteq$.
This reversal of the set inclusion of course does not hold
for the inflationarity notion.

Let us discuss now briefly the functions used
in our considerations. In the preceding sections we clarified 
which of their properties account for specific 
properties of the studied algorithms.  It is tempting then
to confine one's attention to closures, i.e., functions that are
inflationary, monotonic and itempotent.  The importance of closures 
for concurrent constraint programming was recognized by Saraswat, 
Rinard and Panangaden \citeyear{saraswat-semantic} realized and 
for the study of constraint propagation by Benhamou and Older \citeyear{BO97}. 

However, as shown in Apt \citeyear{Apt99b}, some known local consistency
notions are characterized as common fixpoints of functions that 
in general are not itempotent.  Therefore when studying constraint propagation 
in full generality it is preferrable not to limit one's attention to 
closures.  On the other hand, in the sections that follow we only study 
notions of local consistency that are characterized by means of closures.  
Therefore, from now on the closures will be prominently present in our 
exposition.

\section{A Hyper-arc Consistency Algorithm}
\label{sec:hyper-arc-algo}

We begin by considering the notion of hyper-arc consistency of Mohr
and Masini \citeyear{MM88} (we use here the terminology of 
Marriott and Stuckey \citeyear{MS98b}).
The more known notion of arc consistency of Mackworth
\citeyear{mackworth-consistency} is obtained by restricting one's
attention to binary constraints.
Let us recall the definition.

\begin{definition} \mbox{} \\[-6mm]
  \begin{itemize}
  \item 
Consider a constraint $C$ on the variables
$x_1, \LL , x_n$ with the respective domains
$D_1, \LL,  D_n$, that is $C \sse D_1 \times \cdots \times D_n$.
We call $C$ 
{\em hyper-arc consistent\/} if for every $i \in [1..n]$ and $a \in D_i$
there exists $d \in C$ such that $a = d[i]$.

\item We call a CSP {\em hyper-arc consistent\/} if all its
constraints are hyper-arc consistent.
\HB
  \end{itemize}
\end{definition}

Intuitively, a constraint $C$ is hyper-arc consistent
if for every involved domain each element of it participates in a
solution to $C$.

To employ the {\tt CDI} algorithm of Section \ref{sec:compound}
we now make specific choices involving the items (i), (ii) and
(iii) of the previous section.
\II

\NI
Re: (i) Partial orderings with the least elements.

As already mentioned in the previous section, for the function ${\cal
  F}$ we choose the powerset function ${\cal P}$, so for each domain
$D$ we put ${\cal F}(D) := {\cal P}(D)$.

Given a CSP ${\cal P}$ with the sequence $D_1, \LL, D_n$ of the
domains we take the domain reduction ordering with ${\cal P}$ as its
least element. As already noted we can identify this ordering with the
Cartesian product of the partial orderings $({\cal P}(D_i),
\supseteq)$, where $i \in [1..n]$.  The elements of this compound
ordering are thus sequences $(X_1, \LL, X_n)$ of respective subsets of
the domains $D_1, \LL, D_n$ ordered componentwise by the reversed
subset ordering $\supseteq$.
\II

\NI
Re: (ii) Monotonic and inflationary functions with schemes.

Given a constraint $C$ on the variables $y_1, \LL, y_k$ with
respective domains $E_1, \LL, E_k$, we abbreviate for each $j \in
[1..k]$ the set $\C{d[j] \mid d \in C}$ to $\Pi_{j}(C)$.  Thus
$\Pi_{j}(C)$ consists of all $j$-th coordinates of the elements of
$C$.  Consequently, $\Pi_{j}(C)$ is a subset of the domain $E_j$ of
the variable $y_j$.

We now introduce for each $i \in [1..k]$
the following function
$\pi_i$ on ${\cal P}(E_1) \times \cdots \times {\cal P}(E_k)$:

\[
\pi_i(X_1, \LL , X_k) := (X_1, \LL,  X_{i-1}, X'_i, X_{i+1}, \LL, X_k)
\]
where 
\[
X'_i := \Pi_i(C \cap (X_1 \times \cdots \times X_k)).
\]
That is, 
$X'_i = \C{d[i] \mid d \in X_1 \times \cdots \times X_k \mbox { and } d \in C}$.
Each function $\pi_i$ is associated with a specific constraint $C$.
Note that $X'_i \sse X_i$, so each function $\pi_i$ boils down to a
projection on the $i$-th component.  
\II

\NI
Re: (iii) Common fixpoints.

Their use is clarified by the following lemma that also lists the
relevant properties of the functions $\pi_i$ (see  Apt
\citeyear[pages 197 and 202]{Apt99b}).

\begin{lemma}[(Hyper-arc Consistency)] \label{lem:hyper-arc}
 \mbox{} \\[-6mm]
  \begin{enumerate}\smallromani
  \item A CSP $\p{{\cal C}}{x_1 \in D_1, \LL, x_n \in D_n}$
is hyper-arc consistent iff $(D_1, \LL, D_n)$ is a
common fixpoint of all functions $\pi^{+}_i$ associated 
with the constraints from ${\cal C}$.

  \item
Each projection function $\pi_i$ associated with a constraint $C$ is 
a closure w.r.t. the componentwise  ordering $\supseteq$.
\end{enumerate}
\end{lemma}

By taking into account only the binary constraints we obtain an
analogous characterization of arc consistency.  The functions $\pi_1$
and $\pi_2$ can then be defined more directly as follows:
\[
\pi_{1}(X,Y) := (X',Y),
\]
where
$X':= \C{a \in X \mid \te \: b \in Y \: (a,b) \in C}$,
and
\[
\pi_2(X, Y) := (X,Y'),
\]
where $Y' := \C{b \in Y \mid \te a \in X \: (a,b) \in C}$.

Fix now a CSP ${\cal P}$. By instantiating the {\tt CDI} algorithm with 
\[
F_0 := \C{f \mid f \mbox{ is a $\pi_i$ function associated with a
constraint of ${\cal P}$}}
\] 
and with each $\bot_i$ equal to $D_i$ we get the {\tt
HYPER-ARC} algorithm that enjoys following properties.
 
\begin{theorem}[({\tt HYPER-ARC} Algorithm)] \label{thm:hyper-arc}
Consider a CSP
${\cal P} := \p{{\cal C}}{x_1 \in D_1, \LL, x_n \in D_n}$,
where each $D_i$ is finite.

The {\tt HYPER-ARC} algorithm always terminates.
Let ${\cal P'}$ be the CSP determined by ${\cal P}$ and
the sequence of the domains $D'_1, \LL,  D'_n$ 
computed in $d$. Then
\begin{enumerate}\smallromani

\item ${\cal P'}$ is the $\sqsubseteq_d$-least CSP that is hyper-arc consistent,

\item ${\cal P'}$ is equivalent to ${\cal P}$.
\end{enumerate}
\end{theorem}

Due to the definition of the $\sqsubseteq_d$ ordering the item $(i)$ can be
rephrased as follows.  Consider all hyper-arc consistent CSP's that are of
the form $\p{{\cal C'}}{x_1 \in D'_1, \LL, x_n \in D'_n}$ where $D'_i
\sse D_i$ for $i \in [1..n]$ and the constraints in ${\cal C'}$ are
the restrictions of the constraints in ${\cal C}$ to the domains
$D'_1, \LL, D'_n$.  Then among these CSP's ${\cal P'}$ has the largest
domains.

\begin{proof}
The termination and $(i)$ are immediate consequences of the counterpart of
the {\tt CD} Corollary \ref{cor:CD} for the {\tt CDI} algorithm and of
the Hyper-arc Consistency Lemma \ref{lem:hyper-arc}.

To prove $(ii)$ note that the final CSP ${\cal P'}$ can be 
obtained by means of repeated applications of the 
projection functions $\pi_i$ starting with the initial CSP ${\cal P}$.
(Conforming to the discussion at the end of Section \ref{sec:from-to}
we view here each such function as a function on CSP's).
As noted in Apt \citeyear[pages 197 and 201]{Apt99b}) each of these functions
transforms a CSP into an equivalent one.
\end{proof}

\section{An Improvement: the {\tt AC-3} Algorithm}
\label{sec:ac3}

In the {\tt HYPER-ARC} algorithm each time a $\pi_i$ function
associated with a constraint $C$ on the variables $y_1, \LL, y_k$ is
applied and modifies its arguments, all projection functions
associated with a constraint that involves the variable $y_i$ are
added to the set $G$.  In this section we show how we can exploit an
information about the commutativity to add less projection functions
to the set $G$.  Recall that in Section \ref{sec:compound} we modified
the notion of commutativity for the case of functions with schemes.

First, it is worthwhile to note that not all pairs of the 
$\pi_i$ and $\pi_j$ functions commute.

\begin{example} \label{exa:noncom} \mbox{} \\
  (i) We consider the case of two binary constraints on the
  same variables.  Consider two variables, $x$ and $y$ with the
  corresponding domains $D_x := \C{a,b}$, $D_y := \C{c,d}$ and two
  constraints on $x,y$: $C_1 := \C{(a,c), (b,d)}$ and $C_2 :=
  \C{(a,d)}$.

  Next, consider the $\pi_1$ function of $C_1$ and the $\pi_2$
  function of $C_2$.  Then applying these functions in one order,
  namely $\pi_2 \pi_1$, to $(D_x, D_y)$ yields $D_x$ unchanged,
  whereas applying them in the other order, $\pi_1 \pi_2$, yields
  $D_x$ equal to $\C{b}$.  
\II

\NI
(ii) Next, we show that the
commutativity can also be violated due to sharing of a single variable.
As an example take the variables $x,y,z$ with the corresponding domains
$D_x := \C{a,b}$,
$D_y := \C{b}$,
$D_z := \C{c,d}$,
and the constraint $C_1 := \C{(a,b)}$ on $x,y$ and
$C_2 := \C{(a,c), (b,d)}$ on $x,z$. 

Consider now the $\pi_1^{+}$ function of $C_1$ and 
the $\pi_2^{+}$ function of $C_2$.
Then applying these functions in
one order, namely $\pi_2^{+} \pi_1^+$,
to $(D_x, D_y, D_z)$ yields $D_z$ equal to $\C{c}$, whereas
applying them in the other order, $\pi_1^{+} \pi_2^+$,
yields $D_z$ unchanged.
\HB
\end{example}

The following lemma clarifies which projection functions do commute.

\begin{lemma}[(Commutativity)] \label{lem:comm}
Consider a CSP and two constraints of it, $C$ on the variables
$y_1, \LL, y_k$ and $E$ on the variables $z_1, \LL, z_{\ell}$.
\begin{enumerate}\smallromani

\item For $i,j \in [1..k]$ the functions  $\pi_i$ and $\pi_j$ 
of the constraint $C$ commute.

\item If the variables $y_i$ and $z_j$ are identical then the functions $\pi_i$
of $C$ and $\pi_j$ of $E$ commute.

\end{enumerate} 
\end{lemma}

\begin{proof}
See Appendix.  
\end{proof}

Fix now a CSP. We derive a modification of the {\tt HYPER-ARC}
algorithm by instantiating this time the {\tt CDC} algorithm. As 
before we use the set of functions
\[
F_0 := \C{f \mid f \mbox{ is a $\pi_i$ function associated with a
constraint of ${\cal P}$}}
\]
and each $\bot_i$ equal to $D_i$.
Additionally we employ the following function {\em Comm\/},
where $\pi_i$ is associated with a constraint $C$
and where $E$ differs from $C$:

\begin{tabbing}
\quad $Comm(\pi_i)$ \=  := \= \C{\pi_j \mid} \kill 
\quad $Comm(\pi_i)$ \>  := \> \C{\pi_j \mid 
\mbox{$i \neq j $ and $\pi_j$ is associated with the constraint $C$}} \\
\quad             \> $\cup$ \> \{$\pi_j \mid$
\mbox{$\pi_j$ is associated with a constraint $E$ and} \\
            \>        \> \mbox{ \ $\quad$ \ the $i$-th variable of $C$ and the $j$-th variable of $E$ coincide\}}.
\end{tabbing}

By virtue of the Commutativity Lemma \ref{lem:comm} 
each set $Comm(g)$ satisfies the assumptions of the Update Theorem
\ref{thm:update}$(ii)$.

By limiting oneself to the set of functions $\pi_1$ and $\pi_2$
associated with the binary constraints,
we obtain an analogous modification of the 
corresponding arc consistency algorithm.

Using now the counterpart of the {\tt CD} Corollary \ref{cor:CD} for
the {\tt CDC} algorithm we conclude that the above algorithm enjoys
the same properties as the {\tt HYPER-ARC} algorithm, that is the
counterpart of the {\tt HYPER-ARC} Algorithm Theorem
\ref{thm:hyper-arc} holds.

Let us clarify now the difference between this algorithm and the {\tt
HYPER-ARC} algorithm when both of them are limited to the binary
constraints.

Assume that the considered CSP is of the form $\p{{\cal C}}{{\cal DE}}$.
We reformulate the above algorithm as follows.
Given a binary relation $R$, we put
\[
R^{T} := \C{(b,a) \mid (a,b) \in R}.
\]

For $F_0$ we now choose the set of the $\pi_1$ functions of the 
constraints or relations from the set
\begin{tabbing}
\qquad \= $S_0$ \= :=     \= \C{C \mid \mbox{$C$ is a binary constraint from ${\cal C}$}} \\
       \>       \> $\cup$ \> \C{C^{T} \mid \mbox{$C$ is a binary constraint from ${\cal C}$}}.
\end{tabbing}

Finally, for each $\pi_1$ function of some $C \in S_0$ on $x,y$ 
we define
\begin{tabbing}
\qquad  $Comm(\pi_1)$ \=  :=  \= \C{\pi_j \mid} \kill
\qquad $Comm(\pi_1)$ \>  :=  \> \C{\mbox{the $\pi_1$ function of $C^{T}$}} \\
\qquad               \> $\cup$ \> \C{f \mid \mbox{$f$ is the $\pi_1$ function of some $E \in S_0$ on
                         $x,z$ where $z \not \equiv y$}}.
\end{tabbing}

Assume now that
\begin{equation}
  \label{eq:atmost}
\mbox{for each pair of variables $x,y$ at most one constraint exists on $x,y$.}  
\end{equation}

Consider now the corresponding instance of the {\tt CDC} algorithm.
By incorporating into it the effect of the functions $\pi_1$ on the
corresponding domains, we obtain the following
algorithm known as the {\tt AC-3} algorithm of Mackworth
\citeyear{mackworth-consistency}.

We assume here that ${\cal DE} := x_1 \in D_1, \LL, x_n \in D_n$.
\II

\NI
{\sc {\tt AC-3\/} Algorithm}

\begin{tabbing}
\= $S_0$ \= := \= \C{C \mid \mbox{$C$ is a binary constraint from ${\cal C}$}} \\
       \>       \> $\cup$ \> \C{C^{T} \mid \mbox{$C$ is a binary constraint from ${\cal C}$}}; \\
\> $S := S_0$; \\ 
\> {\bf while} $S \neq \ES$ {\bf do} \\
\> \qquad choose $C \in S$; suppose $C$ is on $x_i, x_j$; \\
\> \qquad $D_i := \C{a \in D_i \mid \te \: b \in D_j \: (a,b) \in C}$; \\
\> \qquad {\bf if} $D_i$ changed {\bf then} \\
\> \qquad \qquad $S := S \cup \C{ C' \in S_0 \mid C' 
  \mbox{ is on the variables  $y, x_i$ where $y \not \equiv x_j$} }$ \\
\> \qquad {\bf fi}; \\
\> \qquad $S := S - \C{C}$ \\
\> {\bf od}
\end{tabbing}
\III

It is useful to mention that the corresponding reformulation of the
{\tt HYPER-ARC} algorithm for binary constraints differs in the second
assignment to $S$ which is then
\[
S := S \cup \C{ C' \in S_0 \mid C' \mbox{ is on the variables  $y, z$ where $y$ is $x_i$ or $z$ is $x_i$}}.
\]

So we ``capitalized'' here on the commutativity of the corresponding
projection functions $\pi_1$ as follows. First, no constraint or
relation on $x_i, z$ for some $z$ is added to $S$.  Here we exploited
part $(ii)$ of the Commutativity Lemma \ref{lem:comm}.

Second, no constraint or relation on $x_j, x_i$ is added to $S$.  Here
we exploited part $(i)$ of the Commutativity Lemma \ref{lem:comm},
because by assumption (\ref{eq:atmost}) $C^{T}$ is the
only constraint or relation on $x_j, x_i$ and its $\pi_1$ function
coincides with the $\pi_2$ function of $C$.

In case assumption (\ref{eq:atmost})
about the considered CSP is dropped, the resulting algorithm
is somewhat less readable. However, once
we use the following
modified definition of $Comm(\pi_1)$:
\[
Comm(\pi_1) := \C{f \mid \mbox{$f$ is the $\pi_1$ function of some $E \in S_0$ on $x,z$ where $z \not \equiv y$}}
\]
we get an instance of the {\tt CDC}
algorithm which differs from the {\tt AC-3} algorithm in
that the qualification ``where $y \not \equiv x_j$'' is removed from the
definition of the second assignment to the set $S$.

To illustrate the considerations of this section
let us return now to the crossword puzzle introduced in Subsection
\ref{subsec:example}.

As pointed out by Mackworth \citeyear{mackworth-constraint-encylopedia}
this problem can be easily formulated as a CSP as follows. First,
associate with each position $i \in [1..8]$ in the grid 
of Figure \ref{fig:crossword} a variable.
Then associate with each variable the domain that consists of the set
of words that can be used to fill this position. For example,
position 6 needs to be filled with a three letter word, so the domain
of the variable associated with position 6 consists of the above set
of five 3 letter words.

Finally, we define constraints. They deal with the restrictions
arising from the fact that the words that cross share a letter.  For
example, the crossing of the positions 1 and 2 contributes the
following constraint:

\begin{tabbing}
$C_{1, 2}$ := \{\=(HOSES, SAILS), (HOSES, SHEET), (HOSES, STEER),  \\
\>(LASER, SAILS), (LASER, SHEET), (LASER, STEER)\} .
\end{tabbing}
This constraint formalizes the fact that the third letter of position
1 needs to be the same as the first letter of position 2. In total
there are twelve constraints.

Each projection function $\pi_1$ associated with a constraint $C$ or
its transpose $C^T$ corresponds to a crossing, for example (8,2). It
removes impossible values from the current domain of the variable
associated with the first position, here 8.

The above Commutativity Lemma \ref{lem:comm} allows us to conclude
that for any pairwise different $a,b,c \in [1..8]$ the projection
functions $\pi_1$ associated with the crossings $(a,b)$ and $(b,a)$
commute and also the projection functions $\pi_1$ associated with the
crossings $(a,b)$ and $(a,c)$ commute.  This explains why in the {\tt
  AC-3} algorithm applied to this CSP after considering a crossing
$(a,b)$, for example (2,4), neither the crossing (4,2) nor the
crossings (2,7) and (2,8) are added to the set of examined crossings.

To see that the {\tt AC-3} algorithm applied to this CSP 
yields the unique solution depicted in Figure \ref{fig:crossword1} it is sufficient to
observe that this solution viewed as a CSP is arc consistent
and that it is obtained by a specific execution of the {\tt AC-3} algorithm,
in which the crossings are considered in the following order:

\begin{center}
(1,2), (2,1), (1,3), (3,1), (4,2), (2,4), (4,5), (5,4), (4,2), (2,4),  \\
(7,2), (2,7), (7,5), (5,7), (8,2), (2,8),  (8,6), (6,8), (8,2), (2,8).
\end{center}

The desired conclusion now follows by the counterpart of
the  {\tt CD} Corollary \ref{cor:CD}
according to which every execution of the {\tt AC-3} algorithm
yields the same outcome.

\section{A Path Consistency Algorithm}
\label{sec:path-algo}

The notion of path consistency was introduced in Montanari
\citeyear{montanari-networks}.  It is defined for special type of CSP's.
For simplicity we ignore here unary constraints that are usually
present when studying path consistency.

\begin{definition}
  We call a CSP ${\cal P}$ {\em standardized\/} if for each pair
  $x,y$ of its variables there exists exactly one constraint on $x,y$ in
  ${\cal P}$. We denote this constraint by $C_{x,y}$.
\HB
\end{definition}

Every CSP is trivially equivalent to a standardized CSP.  Indeed, it
suffices for each pair $x,y$ of the variables of ${\cal P}$ first to
add the ``universal'' constraint on $x,y$ that consists of the
Cartesian product of the domains of the variables $x$ and $y$ and then
to replace the set of all constraints on $x,y$ by their intersection.

At the cost of some notational overhead our considerations about path
consistency can be generalized in a straightforward way to the case of
CSP's such that for each pair of variables $x,y$ at most one
constraint exists on $x,y$, i.e., to the CSP's that satisfy assumption
(\ref{eq:atmost}).

To simplify the notation given two binary relations $R$ and $S$
we define their composition $\cdot$ by
\[
R \cdot S :=   \C{(a,b) \mid \te c \: ((a,c) \in R, (c,b) \in S)}.
\]

Note that if $R$ is a constraint on the variables $x,y$ and $S$ a
constraint on the variables $y,z$, then $R \cdot S$ is a
constraint on the variables $x,z$.

Given a subsequence $x,y$ of two variables of a standardized CSP 
we introduce a ``supplementary'' relation $C_{y,x}$ defined by
\[
C_{y,x} := C^{T}_{x,y}.
\]

Recall that the relation $C^{T}$ was introduced in the previous section.
The supplementary relations are not parts of the considered CSP as
none of them is defined on a subsequence of its variables, but they
allow us a more compact presentation. We now introduce the following notion.

\begin{definition} 
We call a standardized CSP {\em path consistent\/}
if for each subset $\C{x,y,z}$ of its variables we have
\[
C_{x,z} \sse C_{x,y} \cdot C_{y,z}.
\]
\HB
\end{definition}

In other words, a standardized CSP is path consistent if
for each subset $\C{x,y,z}$ of its variables the following
holds: 
\begin{center}
if $(a,c) \in C_{x,z}$, then
there exists $b$ such that $(a,b) \in C_{x,y}$ and $(b,c) \in C_{y,z}$.
\end{center}

To employ the {\tt CDI} algorithm of Section \ref{sec:compound}
we again make specific choices involving the items (i), (ii) and
(iii) of Section \ref{sec:from-to}.
First, we provide an alternative characterization of path consistency.

Note that in the above definition we used the relations of the form $C_{u,v}$
for any {\em subset\/} $\C{u,v}$ of the considered sequence of
variables. If $u,v$ is not a {\em subsequence\/} of the original
sequence of variables, then $C_{u,v}$ is a supplementary relation
that is not a constraint of the original CSP. At the expense of some
redundancy we can rewrite the above definition so that only the
constraints of the considered CSP are
involved.  This is the contents of the following simple observation
that will be useful in a moment.

\begin{note}[(Alternative Path Consistency)] \label{not:path-alternative}
A standardized CSP is path consistent iff
for each subsequence $x,y,z$ of its variables we have

\[
C_{x,y} \sse C_{x,z} \cdot C^{T}_{y,z},
\]

\[
C_{x,z} \sse C_{x,y} \cdot C_{y,z},
\]

\[
C_{y,z} \sse C^{T}_{x,y} \cdot C_{x,z}.
\]

\end{note}

\begin{figure}[htbp]
  \begin{center}
\input{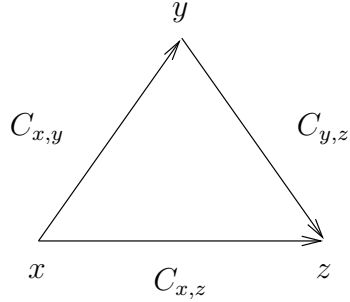}
    \caption{Three relations on three variables}
    \label{fig:path}
  \end{center}
\end{figure}

Figure \ref{fig:path} clarifies this observation. For instance,
an indirect path from $x$ to $y$ via $z$ requires the reversal
of the arc $(y,z)$. This translates to the first formula.

Now, to study path consistency, given a standardized CSP 
${\cal P} := \p{{C_1, \LL, C_k}}{{\cal DE}}$
we take the constraint reduction ordering of of Section \ref{sec:from-to}
with ${\cal P}$ as the least element and with the powerset function 
as the function ${\cal F}$.
So, as already noted in Section \ref{sec:from-to}
we can identify this ordering with the
Cartesian product of the partial orderings $({\cal P}(C_i),
\supseteq)$, where $i \in [1..k]$.  The elements of this compound
ordering are thus sequences $(X_1, \LL, X_k)$ of respective subsets of
the constraints $C_1, \LL, C_k$ ordered componentwise by the reversed
subset ordering $\supseteq$.

Next, given a subsequence $x,y,z$ of the variables of ${\cal P}$ we 
introduce three functions on ${\cal P}(C_{x,y}) \times {\cal
  P}(C_{x,z}) \times {\cal P}(C_{y,z})$:

\[
f^{z}_{x,y}(P,Q,R) := (P',Q,R),
\]
where $P' := P \cap Q \cdot R^T$,
\[
f^{y}_{x,z}(P,Q,R) := (P,Q',R),
\]
where $Q' := Q \cap P \cdot R$,
and
\[
f^{x}_{y,z}(P,Q,R) := (P,Q,R'),
\]
where $R' := R \cap P^T \cdot Q$.

In what follows, when using a function $f^{z}_{x,y}$ we implicitly
assume that the variables $x,y,z$ are pairwise different
and that $x,y$ is a subsequence of the variable of the considered CSP.

Finally, we relate the notion of path consistency to the
common fixpoints of the above defined functions.
This leads us to the following counterpart of the Hyper-arc
Consistency Lemma \ref{lem:hyper-arc}.

\begin{lemma}[(Path Consistency)]\label{lem:path}
 \mbox{} \\[-6mm]
  \begin{enumerate}\smallromani
  \item 
A standardized CSP $\p{{C_1, \LL, C_k}}{{\cal DE}}$
is path consistent iff $(C_1, \LL, C_k)$ is a common fixpoint of all
functions $(f^{z}_{x,y})^{+}$, $(f^{y}_{x,z})^{+}$ and
$(f^{x}_{y,z})^{+}$ associated with
the subsequences $x,y,z$ of its variables.

\item
The functions  $f^{z}_{x,y}$, $f^{y}_{x,z}$ and $f^{x}_{y,z}$ are
closures w.r.t. the componentwise  ordering $\supseteq$.
\end{enumerate}
\end{lemma}
\begin{proof}
$(i)$ is a direct consequence of the Alternative Path
Consistency Note \ref{not:path-alternative}.
The proof of $(ii)$ is straightforward. These properties of the 
functions  $f^{z}_{x,y}$, $f^{y}_{x,z}$ and $f^{x}_{y,z}$ were already 
mentioned in  Apt \citeyear[page 193]{Apt99b}.
\end{proof}

We now instantiate the {\tt CDI} algorithm with the set of functions
\[
F_0 := \C{f \mid x,y,z 
\mbox{ is a subsequence of the variables of ${\cal P}$ and
$f \in \C{f^{z}_{x,y}, f^{y}_{x,z}, f^{x}_{y,z}}$}},
\]
$n := k$ and each $\bot_i$ equal to $C_i$.

Call the resulting algorithm the {\tt PATH} algorithm.
It enjoys the following properties.
 
\begin{theorem}[({\tt PATH} Algorithm)] \label{thm:path-algo}
Consider a standardized CSP \\
${\cal P} := \p{{C_1, \LL, C_k}}{{\cal DE}}$. 
Assume that each constraint $C_i$ is finite. 

The {\tt PATH} algorithm always terminates.  Let 
${\cal P'} := \p{{C'_1, \LL, C'_k}}{{\cal DE}}$, where
the sequence of the
constraints $C'_1, \LL, C'_k$ is computed in $d$. Then

\begin{enumerate}\smallromani

\item ${\cal P'}$ is the $\sqsubseteq_c$-least CSP 
that is path consistent,

\item ${\cal P'}$ is equivalent to ${\cal P}$.

\end{enumerate}
\end{theorem}

As in the case of the {\tt HYPER-ARC} Algorithm Theorem \ref{thm:hyper-arc}
the item  $(i)$
can be rephrased as follows.  Consider all path
consistent CSP's that are of the form $\p{{C'_1, \LL, C'_k}}{{\cal DE}}$
where $C'_i \sse C_i$ for $i \in [1..k]$.  Then among them
${\cal P'}$ has the largest constraints.
\III

\NI
\begin{proof}
The proof is analogous to that of the {\tt HYPER-ARC} 
Algorithm Theorem \ref{thm:hyper-arc}.
The termination and $(i)$ are immediate consequences of the counterpart of
the {\tt CD} Corollary \ref{cor:CD} for the {\tt CDI} algorithm and of
the Path Consistency Lemma \ref{lem:path}.

To prove  $(ii)$ we now note that the final CSP ${\cal P'}$ can be 
obtained by means of repeated applications of the 
functions  $f^{z}_{x,y}$, $f^{y}_{x,z}$ and $f^{x}_{y,z}$
starting with the initial CSP ${\cal P}$.
(Conforming to the discussion at the end of Section \ref{sec:from-to}
we view here each such function as a function on CSP's). 
As noted in  Apt \citeyear[pages 193 and 195]{Apt99b}) each of these functions
transforms a CSP into an equivalent one.
\end{proof}

\section{An Improvement: the {\tt PC-2} Algorithm}
\label{sec:pc2}

In the {\tt PATH} algorithm each time a $f^{z}_{x,y}$ function is
applied and modifies its arguments, all functions associated with a
triplet of variables including $x$ and $y$ are added to the set $G$.
We now show how we can add less functions by taking into account the
commutativity information.
To this end we establish the following lemma.

\begin{lemma}[(Commutativity)] \label{lem:comm-path}
  Consider a standardized CSP involving among others the variables
  $x,y,z,u$.  Then the functions $f^{z}_{x,y}$ and $f^{u}_{x,y}$
  commute.
\end{lemma}

In other words, two functions with the same pair of
variables as a subscript commute.

\begin{proof}[(Sketch)]
The following intuitive argument may help to understand the
more formal justification given in Appendix.  First, both
considered functions have three arguments but 
share precisely one argument, the one from ${\cal P}(C_{x,y})$,
and modify only this shared argument.  Second, both functions
are defined in terms of the set-theoretic intersection operation
``$\cap$'' applied to two, unchanged, arguments. This
yields commutativity since ``$\cap$'' is commutative.
\end{proof}

Fix now a standardized CSP ${\cal P}$. 
We instantiate the {\tt CDC} algorithm with the same set of
functions $F_0$ as in Section \ref{sec:path-algo}.
Additionally, we use the following function $Comm$:

\[
Comm(f^{z}_{x,y}) = \C{f^{u}_{x,y} \mid u \not\in \C{x,y,z}}.
\]

Thus for each function $g$ the set $Comm(g)$ contains precisely $m-3$
elements, where $m$ is the number of variables of the considered
CSP. This quantifies the maximal ``gain'' obtained by using the commutativity
information: at each ``update'' stage of the corresponding instance of
the {\tt CDC} algorithm we add up to $m-3$ less elements than in the case of
the corresponding instance of the {\tt CDI} algorithm considered in
the previous section.  

By virtue of the Commutativity Lemma \ref{lem:comm-path} each set
$Comm(g)$ satisfies the assumptions of the Update Theorem
\ref{thm:update}$(ii)$.  We conclude that the above instance of the
{\tt CDC} algorithm enjoys the same properties as the original {\tt
PATH} algorithm, that is the counterpart of the {\tt PATH} Algorithm
Theorem \ref{thm:path-algo} holds.  To make this modification of the
{\tt PATH} algorithm easier to understand we proceed as follows.

Below we write  $x \prec y$ to indicate that
$x,y$ is a subsequence of the variables of the CSP ${\cal P}$.
Each function of the form $f^{u}_{x,y}$ where $x \prec y$ and $u
\not\in \C{x,y}$ can be identified with the sequence $x,u,y$ of the
variables. (Note that the ``relative'' position of $u$ w.r.t. $x$ and
$y$ is not fixed, so $x,u,y$ does not have to be a subsequence of the
variables of ${\cal P}$.)
This allows us to identify the set of functions $F_0$ with the set
\[
V_0 := \C{(x,u,y) \mid x \prec y, u \not\in \C{x,y}}.
\]

Next, assuming that $x \prec y$,
we introduce the following set of triples of different variables of ${\cal P}$:

\begin{tabbing}
\qquad $V_{x,y}$ \= :=     \= \C{(x,y,u) \mid x \prec u} $\cup$ \C{(y,x,u) \mid y \prec u} \\
                 \> $\cup$ \> \C{(u,x,y) \mid u \prec y} $\cup$ \C{(u,y,x) \mid u \prec x}.
\end{tabbing}

Informally, $V_{x,y}$ is the subset of $V_0$ that consists of the triples
that begin or end with either $x,y$ or $y,x$.
This corresponds to the set of functions in one of the following forms:
$f^{y}_{x,u}, f^{x}_{y,u}, f^{x}_{u,y}$ and $f^{y}_{u,x}$.

The above instance of the {\tt CDC} algorithm then becomes
the following {\tt PC-2} algorithm of Mackworth
\citeyear{mackworth-consistency}. Here
initially $E_{x,y} = C_{x,y}$.
\II

\NI
{\sc {\tt PC-2\/} Algorithm}

\begin{tabbing}
\= $V_0$ := $\C{(x,u,y) \mid x \prec y, u \not\in \C{x,y}}$; \\
\> $V := V_0$; \\ 
\> {\bf while} $V \neq \ES$ {\bf do} \\
\> \qquad choose $p \in V$; suppose $p = (x,u,y)$; \\
\> \qquad apply  $f^{u}_{x,y}$ to its current domains; \\
\> \qquad {\bf if} $E_{x,y}$ changed {\bf then} \\
\> \qquad \qquad $V := V \cup V_{x,y}$; \\
\> \qquad {\bf fi}; \\
\> \qquad $V := V - \C{p}$ \\
\> {\bf od}
\end{tabbing}

Here the phrase ``apply  $f^{u}_{x,y}$ to its current domains''
can be made more precise if the ``relative'' position of $u$ w.r.t. $x$ and
$y$ is known. Suppose for instance that $u$ is ``before'' $x$ and $y$.
Then $f^{u}_{x,y}$ is defined on 
${\cal P}(C_{u,x}) \times {\cal P}(C_{u,y}) \times {\cal P}(C_{x,y})$ by
\[
f^{u}_{x,y}(E_{u,x},E_{u,y},E_{x,y}) := (E_{u,x},E_{u,y}, E_{x,y} \cap E_{u,x}^T \cdot E_{u,y}),
\]
so the above phrase ``apply  $f^{u}_{x,y}$ to its current domains''
can be replaced by the assignment
\[
E_{x,y} := E_{x,y} \cap E_{u,x}^T \cdot E_{u,y}.
\]
Analogously for the other two possibilities.

The difference between the {\tt PC-2\/} algorithm and the corresponding
representation of the {\tt PATH} algorithm lies in the way
the modification of the set $V$ is carried out. In the case
of the {\tt PATH} algorithm the second assignment to $V$ is
\[
V := V \cup V_{x,y} \cup \C{(x,u,y) \mid u \not\in \C{x,y}}.
\]

\section{Simple Iteration Algorithms}
\label{sec:simple-ite}

Let us return now to the framework of Section \ref{sec:gen-ite}.  We
analyze here when the {\bf while} loop of the {\sc Generic Iteration
Algorithm} {\tt GI} can be replaced by a {\bf for} loop.  First, we
weaken the notion of commutativity as follows.

\begin{definition}
Consider a partial ordering  $(D, \po )$ and functions
$f$ and $g$ on $D$.
We say that  $f$ {\em semi-commutes with\/} $g$ ({\em w.r.t. $\po$\/}) if 
$f (g (x)) \po g (f (x))$ for all $x$.
\HB  
\end{definition}

The following lemma provides an answer to the  question just posed.
Here and elsewhere we omit brackets when writing repeated applications
of functions to an argument.

\begin{lemma}[(Simple Iteration)] \label{lem:semi-com}            
  Consider a partial ordering $(D, \po )$ with the least element
  $\bot$. Let $F := f_1, \LL, f_k$ be a finite sequence of closures on
  $(D, \po )$.  Suppose that $f_i$ semi-commutes with $f_j$ for $i >
  j$, that is,
\begin{equation}
f_i f_j(x) \po f_j f_i(x) \mbox{ for $i >j$ and for all $x$.}
\label{equ:semi-com}
\end{equation}
Then $f_1 f_2 \LL f_k(\bot)$ is the least common fixpoint
of the functions from $F$.  
\end{lemma}
\begin{proof}
We prove first that for $i \in [1..k]$ we have
\[
f_i f_1 f_2 \LL f_k(\bot) \po f_1 f_2 \LL f_k(\bot).
\]
Indeed, by the assumption (\ref{equ:semi-com}) we have the 
following string of inclusions,
where the last one is due to the idempotence of the considered functions:
\[
f_i f_1 f_2 \LL f_k(\bot) \po f_1 f_i f_2 \LL f_k(\bot) \po \LL \po
f_1 f_2 \LL f_i f_i \LL f_k(\bot) \po f_1 f_2 \LL f_k(\bot).
\]

Additionally, by the inflationarity of the considered functions,
we also have for $i \in [1..k]$
\[
f_1 f_2 \LL f_k(\bot) \po f_i f_1 f_2 \LL f_k(\bot).
\]

So $f_1 f_2 \LL f_k(\bot)$ is a common fixpoint of the functions from
$F$.  This means that any iteration of $F$ that starts with $\bot$,
$f_k(\bot)$, $f_{k-1} f_k(\bot), \LL, f_1 f_2 \LL f_k(\bot)$
eventually stabilizes at $f_1 f_2 \LL f_k(\bot)$. By the Stabilization
Lemma \ref{lem:stabilization} we get the desired conclusion.  
\end{proof}

The above lemma provides us with a simple way of computing the least
common fixpoint of a finite set of functions that satisfy the
assumptions of this lemma, in particular condition
(\ref{equ:semi-com}). Namely, it suffices to order these functions in
an appropriate way and then to apply each of them just once, starting
with the argument $\bot$.

The following algorithm is a counterpart of the {\tt GI} algorithm.
We assume in it that condition (\ref{equ:semi-com}) holds for
the sequence of functions $f_1, \LL, f_k$.
\II

\NI
{\sc Simple Iteration Algorithm ({\tt SI})}
\begin{tabbing}
\= $d := \bot$; \\
\> {\bf for} $i := k$ {\bf to} $1$ {\bf by} $-1$ {\bf do} \\
\> \qquad $d := f_{i}(d)$ \\
\> {\bf od} 
\end{tabbing}

The following immediate consequence of the Simple Iteration Lemma
\ref{lem:semi-com} is a counterpart of the {\tt GI} Corollary
\ref{cor:GI}.

\begin{corollary}[({\tt SI})] \label{cor:SI}
  Suppose that $(D, \po )$ is a partial ordering with the least
  element $\bot$. Let $F := f_1, \LL, f_k$ be a finite sequence of
  closures on $(D, \po )$ such that (\ref{equ:semi-com}) holds. Then
  the {\tt SI} algorithm terminates and computes in $d$ the least
  common fixpoint of the functions from $F$.
\end{corollary}

Note that in contrast to the {\tt GI} Corollary \ref{cor:GI} we do not
require here that the partial ordering is finite.  We can view the
{\tt SI} algorithm as a specialization of the {\tt GI} algorithm of
Section \ref{sec:gen-ite} in which the elements of the set of
functions $G$ are selected in a specific way and in which the $update$
function always yields the empty set.

In Section \ref{sec:compound} we refined the {\tt GI} algorithm for
the case of compound domains. An analogous refinement of the {\tt SI}
algorithm is straightforward and omitted.  In the next two sections we
show how we can use this refinement of the {\tt SI} algorithm to
derive two well-known constraint propagation algorithms.

\section{{\tt DAC}: a Directional Arc Consistency Algorithm}
\label{sec:directional-arc-algo}

We consider here the notion of directional arc consistency of
Dechter and Pearl \citeyear{dechter88}.
Let us recall the definition. 

\begin{definition}
Assume a linear ordering $\prec$ on the considered variables.

  \begin{itemize}
  \item 
Consider a binary constraint $C$ on the variables $x, y$ with the domains 
$D_x$ and $D_y$. We call $C$
{\em directionally arc consistent  w.r.t. $\prec$\/} if
\begin{itemize}

\item
$\fa a \in D_x \te b \in D_y \: (a,b) \in C$ provided $x \prec y$,

\item
$\fa b \in D_y \te a \in D_x \: (a,b) \in C$ provided $y \prec x$.

\end{itemize}

So out of these two conditions on $C$ exactly one needs to be checked.
\item We call a CSP {\em directionally arc consistent  w.r.t. $\prec$\/} if all its
binary constraints are directionally arc consistent  w.r.t. $\prec$.
\HB
  \end{itemize}
\end{definition}

To derive an algorithm that achieves this local consistency notion we
first characterize it in terms of fixpoints.  To this end, given a
${\cal P}$ and a linear ordering $\prec$ on its variables, we rather
reason in terms of the equivalent CSP ${\cal P}_{\prec}$ obtained from
${\cal P}$ by reordering its variables along $\prec$ so that each
constraint in ${\cal P}_{\prec}$ is on a sequence of variables $x_1,
\LL, x_n$ such that $x_1 \prec x_2 \prec \LL \prec x_n$.

The following simple characterization holds.

\begin{lemma}[(Directional Arc Consistency)] \label{lem:darc} 
  Consider a CSP ${\cal P}$ with a linear ordering $\prec$ on its
  variables.  Let ${\cal P}_{\prec} := \p{{\cal C}}{x_1 \in D_1, \LL,
    x_n \in D_n}$.  Then ${\cal P}$ is directionally arc consistent
  w.r.t. $\prec$ iff $(D_1, \LL, D_n)$ is a common fixpoint of the
  functions $\pi^{+}_1$ associated with the binary constraints from
  ${\cal P}_{\prec}$.
\end{lemma}

We now instantiate in an appropriate way the {\tt SI} algorithm for
compound domains with all the $\pi_1$ functions associated with the
binary constraints from ${\cal P}_{\prec}$.  In this way we obtain an
algorithm that achieves for ${\cal P}$ directional arc consistency
w.r.t. $\prec$.  First, we adjust the definition of semi-commutativity
to functions with different schemes.
To this end consider a sequence of partial orderings $(D_1, \po_1),
\LL , (D_n, \po_n)$ and their Cartesian product $(D, \po )$.  Take
two functions, $f$ with scheme $s$ and $g$ with scheme $t$. We say that $f$
{\em semi-commutes with \/} $g$ ({\em w.r.t. $\po$\/}) if $f^+$ 
semi-commutes with $g^+$ w.r.t. $\po$, that is if
\[
f^+ (g^+ (d)) \po g^+ (f^+ (d))
\]
for all $d \in D$.

The following lemma is crucial.
To enhance the readability, we replace here the
irrelevant variables by $\_$.
\begin{lemma}[(Semi-commutativity)] \label{lem:semi}
Consider a CSP and two binary constraints of it,
$C_1$ on $\_,z$ and $C_2$ on $\_,y$, where $y \preceq z$.

Then the $\pi_1$ function of $C_1$ semi-commutes with
the $\pi_1$ function of $C_2$
w.r.t. the componentwise  ordering $\supseteq$.
\end{lemma}

\begin{proof}
See Appendix.  
\end{proof}

To be able to apply this lemma we order appropriately the $\pi_1$
functions of the binary constraints of ${\cal P}_{\prec}$.  Namely,
given two $\pi_1$ functions, $f$ associated with a constraint on
$\_,z$ and $g$ associated with a constraint on $\_,y$, we put $f$
before $g$ if $y \prec z$. Then by virtue of this lemma and the
Commutativity Lemma \ref{lem:comm}(ii) if the function $f$ precedes
the function $g$, then $f$ semi-commutes with $g$ w.r.t. the
componentwise ordering $\supseteq$. 

Observe that we leave here unspecified the order between two $\pi_1$
functions, one associated with a constraint on $x,z$ and another with
a constraint on $y,z$, for some variables $x,y,z$. Note that if $x$
and $y$ coincide then the semi-commutativity is indeed a consequence
of the Commutativity Lemma \ref{lem:comm}(ii).

We instantiate now the refinement of the {\tt SI} algorithm for the
compound domains by the above-defined sequence of the $\pi_1$ functions 
and each $\bot_i$ equal to the domain $D_i$ of the variable $x_i$.
In this way we obtain the following algorithm, where 
the sequence of functions is $f_1, \LL, f_k$.
\II

\NI
{\sc Directional Arc Consistency Algorithm ({\tt DARC})}
\begin{tabbing}
\= $d := (D_1, \LL, D_n)$; \\
\> {\bf for} $j := k$ {\bf to} $1$ {\bf by} $-1$ {\bf do} \\
\> \qquad suppose $f_j$ is with scheme $s$; \\
\> \qquad $d[s] := f_{j}(d[s])$ \\
\> {\bf od}
\end{tabbing}
\NI

This algorithm enjoys the following properties.

\begin{theorem}[({\tt DARC} Algorithm)] \label{thm:darc-algo}
Consider a CSP ${\cal P}$ with a linear ordering $\prec$ on its
variables.  Let 
${\cal P}_{\prec} := \p{{\cal C}}{x_1 \in D_1, \LL, x_n \in D_n}$. 

The {\tt DARC} algorithm always terminates.  
Let ${\cal P'}$ be the CSP determined by ${\cal P}_{\prec}$ and
the sequence of the domains $D'_1, \LL,  D'_n$ 
computed in $d$. Then
\begin{enumerate}\smallromani

\item ${\cal P'}$ is the $\sqsubseteq_d$-least CSP in
$\C{{\cal P}_1 \mid {\cal P}_{\prec} \sqsubseteq_d {\cal P}_1}$
that is directionally arc consistent w.r.t. $\prec$,

\item ${\cal P'}$ is equivalent to ${\cal P}$.
\end{enumerate}
\end{theorem}
\begin{proof}
The termination is obvious. $(i)$ is an immediate consequences of the
counterpart of the {\tt SI} Corollary \ref{cor:SI} for the {\tt SI}
algorithm refined for the compound domains and of the Directional Arc
Consistency Lemma \ref{lem:darc}.

The proof of $(ii)$ is analogous to that of the {\tt HYPER-ARC} 
Algorithm Theorem \ref{thm:hyper-arc}$(ii)$.
\end{proof}

Note that in contrast to the {\tt HYPER-ARC} Algorithm Theorem
\ref{thm:hyper-arc} we do not need to assume here that each domain is
finite.

Assume now that the original CSP ${\cal P}$ is standardized, i.e., for
each pair of its variables $x,y$ precisely one constraint on $x,y$
exists. The same holds then for ${\cal P}_{\prec}$.  
We now specialize the {\tt DARC} algorithm by ordering the
$\pi_1$ functions in a deterministic way.
Suppose that ${\cal P}_{\prec} := \p{{\cal C}}{x_1 \in D_1, \LL, x_n \in D_n}$.
Denote the unique constraint of ${\cal P}_{\prec}$ on $x_i, x_j$ by
$C_{i,j}$.  

Order now these constraints as follows:
\[
C_{1,n}, C_{2,n}, \LL, C_{n-1,n}, 
C_{2,n-1}, \LL, C_{n-2,n-1}, 
\LL, 
C_{1,2}.
\]
That is, the constraint $C_{i', j'}$ precedes the constraint $C_{i''
  ,j''}$ if the pair $(j'',i')$ lexicographically precedes the pair
$(j',i'')$.  Take now the $\pi_1$ functions of these constraints
ordered in the same way as their constraints. 

The above {\tt DARC} algorithm can then be rewritten as the following
double {\bf for} loop.  The resulting algorithm is known as the {\tt
  DAC} algorithm of Dechter and Pearl \citeyear{dechter88}.  
\II

\NI
\begin{tabbing}
\= {\bf for} $j := n$ {\bf to} $2$ {\bf by} $-1$ {\bf do} \\
\> \qquad {\bf for} $i := 1$ {\bf to} $j-1$ {\bf do} \\
\> \qquad \qquad $D_i := \C{a \in D_i \mid \te \: b \in D_j \: (a,b) \in C_{i,j}}$ \\
\> \qquad {\bf od} \\
\> {\bf od} 
\end{tabbing}

\section{{\tt DPC}: a Directional Path Consistency Algorithm}
\label{sec:directional-path-algo}

In this section we deal with the notion of directional path
consistency defined in Dechter and Pearl \citeyear{dechter88}.  
Let us recall the definition.

\begin{definition} 
Assume a linear ordering $\prec$ on the considered variables.
We call a standardized CSP {\em directionally path consistent w.r.t. $\prec$\/}
if for each subset $\C{x,y,z}$ of its variables we have
\[
C_{x,z} \sse C_{x,y} \cdot C_{y,z} \mbox{ provided $x,z \prec y$}.
\]
\HB
\end{definition}

This definition relies on the supplementary relations
because the ordering $\prec$ may differ from the original ordering of the
variables. For example, in the original ordering $z$ can precede $x$.
In this case $C_{z,x}$ and not $C_{x,z}$ is a constraint of the CSP
under consideration.

But just as in the case of path consistency we can rewrite
this definition using the original constraints only.
In fact, we have the following analogue of the 
Alternative Path Consistency Note \ref{not:path-alternative}.

\begin{note}[(Alternative Directional Path Consistency)] 
\label{not:dir-path-alternative}
$\!$A standar\-dized CSP is directionally path consistent  w.r.t. $\prec$ iff
for each subsequence $x,y,z$ of its variables we have

\[
C_{x,y} \sse C_{x,z} \cdot C^{T}_{y,z}
\mbox{ provided $x,y \prec z$},
\]

\[
C_{x,z} \sse C_{x,y} \cdot C_{y,z}
\mbox{ provided $x,z \prec y$},
\]

\[
C_{y,z} \sse C^{T}_{x,y} \cdot C_{x,z}
\mbox{ provided $y,z \prec x$}.
\]
\end{note}

Thus out of the above three inclusions precisely one needs to be checked.

As before we now characterize this local consistency notion in terms
of fixpoints.  To this end, as in the previous section, given a
standardized CSP ${\cal P}$ we rather consider the equivalent CSP
${\cal P}_{\prec}$.  The variables of ${\cal P}_{\prec}$ are ordered
according to $\prec$ and ${\cal P}_{\prec}$ is standardized, as well.

The following counterpart of the
Directional Arc Consistency Lemma \ref{lem:darc}
is a direct consequence of the Alternative Directional Path
Consistency Note \ref{not:dir-path-alternative}.
We use here the functions $f^{z}_{x,y}$ defined in Section 
\ref{sec:path-algo}.

\begin{lemma}[(Directional Path Consistency)] \label{lem:dpath}
  Consider a standardized \\
CSP ${\cal P}$ with a linear ordering $\prec$ on its
  variables.  Let ${{\cal P}_{\prec}} := \p{{C_1, \LL,
      C_k}}{{\cal DE}}$.  Then ${\cal P}$ is directionally path
  consistent w.r.t. $\prec$ iff
$(C_1, \LL, C_k)$ is a common fixpoint of all
functions $(f^{z}_{x,y})^{+}$, where $x \prec y \prec z$.

\end{lemma}

To obtain an algorithm that achieves directional path consistency we
now instantiate in an appropriate way the {\tt SI} algorithm.  To this
end we need the following lemma.

\begin{lemma}[(Semi-commutativity)] \label{lem:semip}
Consider a standardized CSP with a linear ordering $\prec$ on its
variables. Suppose that
$x_1 \prec y_1  \prec z$, $x_2 \prec y_2 \prec u$ and
$u \preceq z$.
Then the function $f^{z}_{x_1 ,y_1}$ semi-commutes with
the function $f^{u}_{x_2, y_2}$
w.r.t. the componentwise  ordering $\supseteq$.
\end{lemma}

\begin{proof}
  See Appendix.
\end{proof}

Consider now a standardized CSP ${\cal P}$ with a linear ordering
$\prec$ on its variables and the corresponding CSP ${{\cal
    P}_{\prec}}$.  To be able to apply the above lemma we order the
$f^{z}_{x,y}$ functions, where $x \prec y \prec z$, as follows.

Assume that $x_1, \LL, x_n$ is the sequence of the variables of
${{\cal P}_{\prec}}$, that is $x_1 \prec x_2 \prec \LL \prec x_n$.
Let for $m \in [3..n]$ the sequence $L_m$ consist of the functions
$f^{x_m}_{x_i, x_j}$, where $i < j < m$, ordered in an arbitrary way.
Consider the sequence resulting from appending the sequences $L_n,
L_{n-1}, \LL, L_3$, in that order. Then by virtue of the
Semi-commutativity Lemma \ref{lem:semip} if the function $f$ precedes
the function $g$, then $f$ semi-commutes with $g$ w.r.t.  the
componentwise ordering $\supseteq$.

We instantiate now the refinement of the {\tt SI} algorithm for the
compound domains by the above-defined sequence of functions $f^{z}_{x
  , y}$ and each $\bot_i$ equal to the constraint $C_i$. This yields
the {\sc Directional Path Consistency Algorithm ({\tt DPATH})} that
apart from the different choice of the constituent partial
orderings is identical to the {\sc Directional Arc Consistency
  Algorithm} {\tt DARC} of the previous section.  Consequently, the
{\tt DPATH} algorithm enjoys analogous properties as the {\tt DARC}
algorithm. They are summarized in the following theorem.

\begin{theorem}[({\tt DPATH} Algorithm)] \label{thm:dpath-algo}
Consider a standardized CSP ${\cal P}$ with a linear ordering $\prec$ on its
variables.  Let
${{\cal P}_{\prec}} := \p{{C_1, \LL, C_k}}{{\cal DE}}$. 

The {\tt DPATH} algorithm always terminates.
Let ${\cal P'} := \p{{C'_1, \LL, C'_k}}{{\cal DE}}$,
where the sequence of the constraints $C'_1, \LL,  C'_k$ is
computed in $d$. Then
\begin{enumerate}\smallromani

\item ${\cal P'}$ is the $\sqsubseteq_c$-least CSP in
$\C{{\cal P}_1 \mid {{\cal P}_{\prec}} \sqsubseteq_d {\cal P}_1}$
that is directionally path consistent w.r.t. $\prec$,

\item ${\cal P'}$ is equivalent to ${\cal P}$.
\end{enumerate}
\end{theorem}

As in the case of the {\tt DARC} Algorithm Theorem \ref{thm:darc-algo}
we do not need to assume here that each domain is finite.

Let us order now each sequence $L_m$ in such a way
that the function $f^{x_m}_{x_{i'} ,
  x_{j'}}$ precedes $f^{x_m}_{x_{i''}, x_{j''}}$ if the pair $(j',i')$
lexicographically precedes the pair$(j'',i'')$. 
Denote the unique constraint of ${\cal P}_{\prec}$ on $x_i, x_j$ by
$C_{i,j}$.  The above {\tt DPATH} algorithm can then be rewritten as
the following triple {\bf for} loop.  The resulting algorithm is known
as the {\tt DPC} algorithm of Dechter and Pearl \citeyear{dechter88}.
\II

\NI
\begin{tabbing}
\= {\bf for} $m := n$ {\bf to} $3$ {\bf by} $-1$ {\bf do} \\
\> \qquad {\bf for} $j := 2$ {\bf to} $m-1$ {\bf do} \\
\>  \qquad \qquad {\bf for} $i := 1$ {\bf to} $j-1$ {\bf do} \\
\>  \qquad \qquad \qquad $C_{i,j} := C_{i,j} \cap C_{i,m} \cdot C^T_{j,m}$ \\
\>  \qquad \qquad {\bf od} \\
\> \qquad {\bf od} \\
\> {\bf od}
\end{tabbing}

\section{Conclusions and Recent Work}

\label{sec:conclusions}

In this article we introduced a general framework for constraint
propagation. It allowed us to present and explain various constraint
propagation algorithms in a uniform way. 
By starting the presentation with generic iteration algorithms
on arbitrary partial orders we clarified the role played 
in the constraint propagation algorithms by the notions of
commutativity and semi-commutativity.  This in turn allowed us to
provide rigorous and uniform correctness proofs of the {\tt AC-3},
{\tt PC-2}, {\tt DAC} and {\tt DPC} algorithms.  

The following table summarizes the results of this paper.
\III

\NI
\begin{tabular}{|l|l|l|l|}
\hline
Local Consistency & Algorithm  & Generic                        & Lemmata Accounting \\
Notion            &            & Algorithm used                 & for Correctness \\
\hline \hline
arc               & {\tt AC-3} & {\tt CDC}                      & Hyper-arc Consistency \ref{lem:hyper-arc}, \\
consistency &            & (Section \ref{sec:compound})   & Commutativity \ref{lem:comm} \\[2mm]
path              & {\tt PC-2} & {\tt CDC}                      & Path Consistency \ref{lem:path}, \\
consistency       &            & (Section \ref{sec:compound})   & Commutativity \ref{lem:comm-path} \\[2mm]
directional arc   & {\tt DAC}  & {\tt SI}                       & Hyper-arc Consistency \ref{lem:hyper-arc}, \\
consistency       &            & (Section \ref{sec:simple-ite}) & Semi-commutativity \ref{lem:semi} \\[2mm]
directional path  & {\tt DPC}  &  {\tt SI}                      & Hyper-arc Consistency \ref{lem:hyper-arc}, \\
consistency       &            & (Section \ref{sec:simple-ite}) & Semi-commutativity \ref{lem:semip} \\
\hline
\end{tabular}
\III

Since the time this paper was submitted for publication the line of
research here presented was extended in a number of ways.  First,
Gennari \citeyear{Gen00a} extended slightly the framework of this
paper and used it to explain the {\tt AC-4} algorithm of Mohr and
Henderson \citeyear{MH86}, the {\tt AC-5} algorithm of Van Hentenryck,
Deville and Teng \citeyear{vanhentenryck-generic}, and the {\tt GAC-4}
algorithm of Mohr and Masini \citeyear{MM88}.  The complication was
that these algorithms operate on some extension of the original CSP.

Then, Bistarelli, Gennari and Rossi \citeyear{BGR00a} studied
constraint propagation algorithms for soft constraints. To this end
they combined the framework of Apt \citeyear{Apt99b} and of this paper
with the one of Bistarelli, Montanari and Rossi \citeyear{BMR97}. The
latter provides a unified model for several classes of ``nonstandard''
constraint satisfaction problems employing the concept of a semiring.
 
Recently Gennari \citeyear{Gen00b} showed how another
modification of the framework here presented can be used to explain
the {\tt PC-4} path consistency algorithm of Han and Lee
\citeyear{HL88} and the {\tt KS} algorithm of \citeyear{cooper:ai:1989} that
can achieve either $k$-consistency or strong $k$-consistency.

We noted already in Apt \citeyear{Apt99b} that using a single
framework for presenting constraint propagation algorithms makes it
easier to automatically derive, verify and compare these algorithms.
In the meantime the work of Monfroy and R{\'{e}}ty \citeyear{MR99}
showed that this approach also allows us to parallelize constraint
propagation algorithms in a simple and uniform way.  This resulted in
a general framework for distributed constraint propagation algorithms.
As a follow up on this work Monfroy \citeyear{Mon00} showed that it is
possible to realize a control-driven coordination-based version of the
generic iteration algorithm. This shows that constraint propagation
can be viewed as the coordination of cooperative agents.

Additionally, as already noted to large extent in Benhamou
\citeyear{Ben96}, such a general framework facilitates the combination
of these algorithms, a property often referred to as ``solver
cooperation''.  For a coordination-based view of solver cooperation
inspired by such a general approach to constraint propagation see
Monfroy and Arbab \citeyear{MA00}.

Let us mention also that Fern{\'{a}}ndez and Hill \citeyear{FH99}
combined the approach of Apt \citeyear{Apt99b} with that of Codognet
and Diaz \citeyear{CD96a} to construct a general framework for solving
interval constraints defined over arbitrary lattices.  Finally, the
generic iteration algorithm {\tt GI} and its specializations can be
used as a template for deriving specific constraint propagation
algorithms in which particular scheduling strategies are employed.
This was done for instance in Monfroy \citeyear{Mon99} for the case of
non-linear constraints on reals where the functions to be scheduled
were divided into two categories: ``weaker'' and ``stronger'' with the
preference for scheduling the weaker functions first.

Currently we investigate whether existing constraint propagation
algorithms could be improved by using the notions of commutativity and
semi-commutativity.

\section*{APPENDIX}

\begin{proof}[of Commutativity Lemma \ref{lem:comm}]

\mbox{} \\
$(i)$ It suffices to notice that for each $k$-tuple $X_1, \LL, X_k$ 
of subsets of the domains of the respective variables we have

\begin{tabbing}
\qquad $\pi_j(\pi_i(X_1, \LL, X_k))$ \= = $(X_1, \LL, X_{i-1}, X'_i, X_{i+1}, \LL, X_{j-1}, X'_j, X_{j+1}, \LL, X_{k})$ \\
                        \> = $\pi_i(\pi_j(X_1, \LL, X_k))$,
\end{tabbing}
where 
\[
X'_i := \Pi_i(C \cap (X_1 \times \cdots \times X_k)), 
\]
\[
X'_j := \Pi_j(C \cap (X_1 \times \cdots \times X_k)), 
\]
and where we assumed that $i <j$.
\II

\NI
$(ii)$ Let the considered CSP be of the form $\p{{\cal C}}{x_1 \in
  D_1, \LL, x_n \in D_n}$.  Assume that some common variable of $y_1,
\LL, y_k$ and $z_1, \LL, z_{\ell}$ is identical to the variable $x_h$.
Further, let $Sol(C,E)$ denote the set of 
$d \in D_1 \times \LL \times D_n$ such that $d[s] \in C$ and $d[t] \in E$,
where $s$ is the scheme of $C$ and $t$ is the scheme of $E$.

Finally, let $f$ denote the $\pi_i$ function of $C$
and $g$  the $\pi_j$ function of $E$.
It is easy to check that for each $n$-tuple 
$X_1, \LL, X_n$  of subsets of $D_1, \LL, D_n$,
respectively, we have
\begin{tabbing}
\qquad $\pi_i^+(\pi_j^+(X_1, \LL, X_n))$ \= = $(X_1, \LL, X_{h-1}, X'_h, X_{h+1}, \LL, X_{n})$ \\
                       \> = $\pi_j^+(\pi_i^+(X_1, \LL, X_n))$,
\end{tabbing}
where
\[
X'_h := \Pi_h(Sol(C,E) \cap (X_1 \times \cdots \times X_n)).
\]
\end{proof}

\begin{proof}[of Commutativity Lemma \ref{lem:comm-path}]
Note first that the ``relative'' positions
of $z$ and of $u$ w.r.t. $x$ and $y$ are not specified. There are in total
three possibilities concerning $z$ and three possibilities concerning
$u$. For instance, $z$ can be ``before'' $x$ , ``between'' $x$ and $y$
or ``after'' $y$.  So we have to consider in total nine cases.

In what follows we limit ourselves to an analysis of three
representative cases. The proof for the remaining six cases is
completely analogous. 
Recall that we write  $x \prec y$ to indicate that
$x,y$ is a subsequence of the variables of ${\cal P}$.
\\[2mm] 
{\em Case 1}. $y \prec z$ and $y \prec u$.

\begin{figure}[htbp]
  \begin{center}
\input{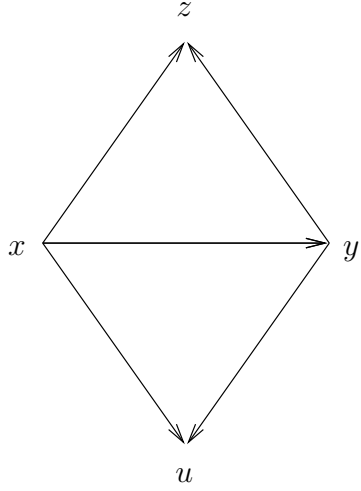}
    \caption{Four variables connected by directed arcs}
    \label{fig:triangles}
  \end{center}
\end{figure}

It helps to visualize these variables as in Figure
\ref{fig:triangles}.  Informally, the functions $f^{z}_{x,y}$ and
$f^{u}_{x,y}$ correspond, respectively, to the upper and lower
triangle in this figure.  The fact that these triangles share an edge
corresponds to the fact that the functions $f^{z}_{x,y}$ and
$f^{u}_{x,y}$ share precisely one argument, the one from ${\cal
  P}(C_{x,y})$.

Ignoring the arguments that do not correspond to the schemes of the
functions $f^{z}_{x,y}$ and $f^{u}_{x,y}$ we can assume that
the functions $(f^{z}_{x,y})^{+}$ and $(f^{u}_{x,y})^{+}$
are both defined on 
\[
{\cal P}(C_{x,y}) \times {\cal P}(C_{x,z}) \times {\cal P}(C_{y,z}) \times 
{\cal P}(C_{x,u}) \times {\cal P}(C_{y,u}).
\]
Each of these functions changes only the first argument.
In fact, for all elements $P,Q,R,U,V$ of, respectively,
${\cal P}(C_{x,y}), {\cal P}(C_{x,z}), {\cal P}(C_{y,z}), {\cal P}(C_{x,u})$
and ${\cal P}(C_{y,u})$, we have
\begin{tabbing}
\qquad $(f^{z}_{x,y})^{+}(f^{u}_{x,y})^{+}(P,Q,R,U,V)$ \= = 
                            $(P \cap U \cdot V^T \cap Q \cdot R^T, Q,R,U,V)$ \\
                       \> = $(P \cap Q \cdot R^T \cap U \cdot V^T, Q,R,U,V)$ \\
                       \> = $(f^{u}_{x,y})^{+}(f^{z}_{x,y})^{+}(P,Q,R,U,V)$.
\end{tabbing}
\II

\NI
{\em Case 2}. $x \prec z \prec y \prec u$.

The intuitive explanation is analogous as in {\em Case 1}. We confine ourselves
to noting that $(f^{z}_{x,y})^{+}$ and $(f^{u}_{x,y})^{+}$ are now defined on
\[
{\cal P}(C_{x,z}) \times {\cal P}(C_{x,y}) \times {\cal P}(C_{z,y}) \times 
{\cal P}(C_{x,u}) \times {\cal P}(C_{y,u})
\]
but each of them changes only the second argument.
In fact, we have

\begin{tabbing}
\qquad $(f^{z}_{x,y})^{+}(f^{u}_{x,y})^{+}(P,Q,R,U,V)$ \= = 
                            $(P, Q \cap U \cdot V^T \cap P \cdot R, R,U,V)$ \\
                       \> = $(P, Q \cap P \cdot R \cap U \cdot V^T,R,U,V)$ \\
                       \> = $(f^{u}_{x,y})^{+}(f^{z}_{x,y})^{+}(P,Q,R,U,V)$.
\end{tabbing}
\II

\NI
{\em Case 3}. $z \prec x$ and $y \prec u$.

In this case the functions $(f^{z}_{x,y})^{+}$ and $(f^{u}_{x,y})^{+}$
are defined on
\[
{\cal P}(C_{z,x}) \times {\cal P}(C_{z,y}) \times {\cal P}(C_{x,y}) \times 
{\cal P}(C_{x,u}) \times {\cal P}(C_{y,u})
\]
but each of them changes only the third argument.
In fact, we have 
\begin{tabbing}
\qquad $(f^{z}_{x,y})^{+}(f^{u}_{x,y})^{+}(P,Q,R,U,V)$ \= = 
                            $(P,Q,R \cap U \cdot V^{T} \cap P^T \cdot Q,U,V)$ \\
                       \> = $(P,Q,R \cap P^T \cdot Q \cap U \cdot V^{T},U,V)$ \\
                       \> = $(f^{u}_{x,y})^{+}(f^{z}_{x,y})^{+}(P,Q,R,U,V)$.
\end{tabbing}
\end{proof}

\begin{proof}[of Semi-commutativity Lemma \ref{lem:semi}]
Suppose that the constraint $C_1$ on the variables $u,z$ and 
the constriant $C_2$ on the variables $x,y$, where $y \preceq z$.
Denote by $f_{u,z}$ the $\pi_1$ function of $C_1$ and by
$f_{x,y}$ the $\pi_1$ function of $C_2$.
The following cases arise. \\[2mm]
{\em Case 1}. $\C{u,z} \cap \C{x,y} = \ES$.

Then the functions $f_{u,z}$ and $f_{x,y}$ commute since their
schemes are disjoint.
\\[2mm] 
{\em Case 2}. $\C{u,z} \cap \C{x,y} \neq \ES$.\\[1mm] 
{\em Subcase 1}. $u = x$.

Then the functions $f_{u,z}$ and $f_{x,y}$ commute by virtue of the 
Commutativity Lemma \ref{lem:comm}$(ii)$. \\[1mm]
{\em Subcase 2}. $u = y$.

 Let the considered CSP be of the form $\p{{\cal C}}{x_1 \in
  D_1, \LL, x_n \in D_n}$.
We can rephrase the claim as follows, where we denote now
$f_{u,z}$ by $f_{y,z}$:
For all $(X_1, \LL, X_n) \in {\cal P}(D_1) \times \cdots \times {\cal P}(D_n)$
we have
\[
f_{y,z}^{+} (f_{x,y}^{+}(X_1, \LL, X_n))  \supseteq 
f_{x,y}^{+}(f_{y,z}^{+}(X_1, \LL, X_n)).
\]

To prove it note first that for some $i,j,k \in [1..n]$ such 
that $i < j < k$ we have
$x = x_i$, $y = x_j$ and $z = x_k$.
We now have
\begin{tabbing}
\qquad $f_{y,z}^{+} (f_{x,y}^{+}(X_1, \LL, X_n))$ \= = $(f_{y,z})^{+}(X_1, \LL, X_{i-1}, X'_i, X_{i+1}, \LL, X_{n})$ \\
    \> = $(X_1, \LL, X_{i-1}, X'_i, X_{i+1}, \LL, X_{j-1}, X'_j, X_{j+1}, \LL, X_{n})$, \\
\end{tabbing}
where
\[
f_{x,y}(X_i, X_j) = (X'_i, X_j)
\]
and 
\[
f_{y,z}(X_j, X_k) = (X'_j, X_k),
\]
whereas

\begin{tabbing}
\qquad $f_{x,y}^{+} (f_{y,z}^{+}(X_1, \LL, X_n))$ \= = $(f_{x,y})^{+}(X_1, \LL, X_{j-1}, X'_j, X_{j+1}, \LL, X_{n})$ \\
    \> = $(X_1, \LL, X_{i-1}, X''_i, X_{i+1}, \LL, X_{j-1}, X'_j, X_{j+1}, \LL, X_{k})$, \\
\end{tabbing}
where
\[
f_{x,y}(X_i, X'_j) = (X''_i, X'_j).
\]

By the Hyper-arc Consistency Lemma \ref{lem:hyper-arc}$(ii)$ each
function $\pi_i$ is inflationary and monotonic w.r.t. the
componentwise ordering $\supseteq$. By the first property applied to
$f_{y,z}$ we have $X_j \supseteq X'_j$, so by the second property
applied to $f_{x,y}$ we have $X'_i \supseteq X''_i$.

This establishes the claim.
\\[1mm] 
{\em Subcase 3}. $z = x$.

This subcase cannot arise since then the variable $z$ precedes 
the variable $y$ whereas by assumption the converse is the case.
\\[1mm]
{\em Subcase 4}. $z = y$.

We can assume by {\em Subcase 1\/} that $u \neq x$. Then 
the functions  $f_{u,z}$ and $f_{x,y}$ commute since
each of them can change only its first component
and this component does not appear in the scheme of the
other function.
\II

This concludes the proof.
\end{proof}

\begin{proof}[of Semi-commutativity Lemma \ref{lem:semip}]
  
  Recall that we assumed that $x_1 \prec y_1 \prec z$, $x_2 \prec y_2
  \prec u$ and $u \preceq z$.  We are supposed to prove that the
  function $f^{z}_{x_1 ,y_1}$ semi-commutes with the function
  $f^{u}_{x_2, y_2}$ w.r.t.  the componentwise ordering $\supseteq$.
The following cases arise.  \\[2mm]
{\em Case 1}. $(x_1, y_1) = (x_2, y_2)$.

In this and other cases by an equality between two pairs
of variables we mean that both the first component variables,
here $x_1$ and $x_2$, and the second component variables,
here $y_1$ and $y_2$, are identical.

In this case the functions $f^{z}_{x_1 ,y_1}$ and $f^{u}_{x_2, y_2}$
commute by virtue of the Commutativity Lemma \ref{lem:comm-path}.
\\[2mm] 
{\em Case 2}. $(x_1, y_1) = (x_2, u)$.

Then $u$ and $z$ differ, since $y_1 \prec z$.
Ignoring the arguments that do not correspond to the schemes of the
functions $f^{z}_{x_1 ,y_1}$ and $f^{u}_{x_2, y_2}$ we can assume that
the functions $(f^{z}_{x_1 ,y_1})^{+}$ and $(f^{u}_{x_2 ,y_2})^{+}$
are both defined on
\[
{\cal P}(C_{x_1 ,y_1}) \times {\cal P}(C_{x_1 ,z}) \times {\cal P}(C_{y_1 ,z}) 
\times {\cal P}(C_{x_2 , y_2}) \times {\cal P}(C_{y_2 ,u}).
\]

The following now holds for
all elements $P,Q,R,U,V$ of, respectively,
${\cal P}(C_{x_1 ,y_1})$, ${\cal P}(C_{x_1 ,z})$, ${\cal P}(C_{y_1 ,z})$,
${\cal P}(C_{x_2 , y_2})$ and ${\cal P}(C_{y_2 ,u})$:

\begin{tabbing}
\qquad $(f^{z}_{x_1 ,y_1})^{+}(f^{u}_{x_2 ,y_2})^{+}(P,Q,R,U,V)$ \= = 
                            $(f^{z}_{x_1 ,y_1})^{+}(P,Q,R, U \cap P \cdot V^T,V)$ \\
                       \> = $(P \cap Q \cdot R^T,R, U \cap P \cdot V^T,V)$ \\
                       \> $\supseteq$ $(P \cap Q \cdot R^T,R, U \cap (P \cap Q \cdot R^T) \cdot V^T,V)$ \\
                       \> = $(f^{u}_{x_2 ,y_2})^{+}(P \cap Q \cdot R^T,Q,R,U,V)$ \\
                       \> = $(f^{u}_{x_2 ,y_2})^{+}(f^{z}_{x_1 ,y_1})^{+}(P,Q,R,U,V)$.
\end{tabbing}

\NI
{\em Case 3}. $(x_1, y_1) = (y_2, u)$.

In this case $u$ and $z$ differ as well, since $y_1 \prec z$.
Again ignoring the arguments that do not correspond to the schemes of the
functions $f^{z}_{x_1 ,y_1}$ and $f^{u}_{x_2, y_2}$ we can assume that
the functions $(f^{z}_{x_1 ,y_1})^{+}$ and $(f^{u}_{x_2 ,y_2})^{+}$
are both defined on

\[
{\cal P}(C_{x_1 ,y_1}) \times {\cal P}(C_{x_1 ,z}) \times {\cal P}(C_{y_1 ,z}) 
\times {\cal P}(C_{x_2 , y_2}) \times {\cal P}(C_{x_2 ,u}).
\]

The following now holds for
all elements $P,Q,R,U,V$ of, respectively,
${\cal P}(C_{x_1 ,y_1})$, ${\cal P}(C_{x_1 ,z})$, ${\cal P}(C_{y_1 ,z})$,
${\cal P}(C_{x_2 , y_2})$ and ${\cal P}(C_{x_2 ,u})$:

\begin{tabbing}
\qquad $(f^{z}_{x_1 ,y_1})^{+}(f^{u}_{x_2 ,y_2})^{+}(P,Q,R,U,V)$ \= = 
                            $(f^{z}_{x_1 ,y_1})^{+}(P,Q,R, U \cap V \cdot P^T,V)$ \\
                       \> = $(P \cap Q \cdot R^T,R, U \cap V \cdot P^T,V)$ \\
                       \> $\supseteq$ $(P \cap Q \cdot R^T,R, U \cap V \cdot (P \cap Q \cdot R^T)^T,V)$ \\
                       \> = $(f^{u}_{x_2 ,y_2})^{+}(P \cap Q \cdot R^T,Q,R,U,V)$ \\
                       \> = $(f^{u}_{x_2, y_2})^{+}(f^{z}_{x_1 ,y_1})^{+}(P,Q,R,U,V)$.
\end{tabbing}

\NI
{\em Case 4}. $(x_1, y_1) \not\in \C{(x_2, y_2), (x_2, u), (y_2, u)}$. 

Then also $(x_2, y_2) \not\in \C{(x_1, y_1), (x_1, z), (y_1, z)}$,
since $(x_2, y_2) \neq (x_1, y_1)$ and $y_2 \neq z$ as
$y_2 \prec u \preceq z$.

Thus the functions $f^{z}_{x_1,y_1}$ and $f^{u}_{x_2,y_2}$ commute
since each of them can change only its first component and this
component does not appear in the scheme of the other function.
\II

This concludes the proof.
\end{proof}
\begin{acks}
Victor Dalmau and Rosella Gennari pointed out to us that in Apt
\citeyear{Apt99c} Assumptions {\bf A} and {\bf B} on page 4 are not
sufficient to establish Theorem 1.  The added now Assumption {\bf C}
was suggested to us by Rosella Gennari.
The referees, the editor Alex Aiken and Eric Monfroy made useful suggestions
concerning the presentation.
\end{acks}

\bibliographystyle{acmtrans}
\bibliography{/ufs/apt/esprit/esprit,/ufs/apt/bib/clpn1,/ufs/apt/bib/pcp,/ufs/apt/bib/99,/ufs/apt/bib/clp2,/ufs/apt/book-ao-2nd/apt,/ufs/apt/book-lp/man1,/ufs/apt/book-lp/man2,/ufs/apt/book-lp/man3,/ufs/apt/book-lp/ref1,/ufs/apt/book-lp/ref2}

\received
Received 11/29/99, Accepted 11/29/00

\endreceived

\end{document}